 \documentclass[11pt]{article}

 \pdfoutput=1

\usepackage{graphicx} % for .epsi formats
\usepackage{amsmath}

\usepackage{enumitem}

\usepackage[francais,english]{babel}

\usepackage{float}
\usepackage{ifthen}
\usepackage{hyperref}

\begin{document}

%-------------------------------------------------------------------

\title{On the definition of a moist-air potential vorticity.}

\author{by Pascal Marquet. {\it M\'et\'eo-France} (pascal.marquet@meteo.fr)}

%\corraddr{Pascal MARQUET, DPr\'evi/Labo, M\'et\'eo-France, 42 av. G. Coriolis, 31057 Toulouse CEDEX 01, France.\\ Web site: http://perso.numericable.fr/$\sim$pmarquet/  ; E-mail: pascal.marquet@meteo.fr }

%\date{\today}
\date{14th of January, 2014}

\maketitle

%-------------------------------------------------------------------

\begin{center}
\noindent
{\em Copy of a paper submitted in August 2012 to the Quarterly Journal of the Royal Meteorological Society (accepted 16 April 2013). With coloured figures.}
\end{center}
\vspace{1mm}

%-------------------------------------------------------------------

\begin{abstract}
A  new potential vorticity is derived by using a specific entropy formulation expressed in terms of a moist-air entropy potential temperature.
The new formulation is compared with  Ertel's version and with others based on virtual and equivalent potential temperatures.
The new potential vorticity is subject to conservative properties ensured by the Second Law applied to the moist-air material derivatives.
It is shown that the upper tropospheric and stratospheric (dry) structures are nearly the same as those obtained with Ertel's component.
Moreover, new structures are observed in the low troposphere, with negative values associated with moist frontal regions.
The negative values are observed in the frontal regions where slantwise convection instabilities may take place, but they are smaller than those observed with the equivalent potential vorticity.
The main purpose  of the article is to diagnose the behaviour of the new potential vorticity from numerical output generated by the ARPEGE NWP model, with the help of isobaric charts and vertical cross-sections.
Two inversion methods are suggested.
The first method could be based on the  invertibility principle verified by the virtual potential vorticity, with a possibility to control and modify separately potential vorticity components  in the (dry) upper and (moist) lower atmospheric levels.
The other method may consist of an inversion process directly applied to the new moist-air entropy potential vorticity, because the negative values and the solenoidal term are smaller than those observed with equivalent potential vorticity, as shown by numerical evaluations.
\end{abstract}

%-------------------------------------------------------------------

%==========================================
\section{Introduction.} % (Section 1)
%==========================================
\label{section_Intro}

The aim of the present article is to extend the definition of the moist version of the potential vorticity defined in Schubert {\it et al.\/} (2001, hereafter referred to as S01), by taking into account the moist entropy formulation derived in Marquet (2011, hereafter referred to as M11).

With the notation introduced in Appendix~A and according to Hoskins {\it et al.\/} (1985, hereafter referred to as H85), the potential vorticity function of any real variable $\psi$  may be defined by
\begin{align}
PV(\psi)  & 
  \: = \; \frac{1}{\rho}\;\boldsymbol{\zeta}_a\: .\: \boldsymbol{\nabla}(\psi)
  \: . \label{def_PV_psi}
\end{align}
It depends on the density $\rho$ and the absolute vorticity $\boldsymbol{\zeta}_a = 2 \: \boldsymbol{\Omega} \: + \:  \boldsymbol{\nabla} \!\times \! \boldsymbol{u}$.
The corresponding $PV$ equation, valid for a hydrostatic atmosphere and for any scalar variable $\psi$, is written as
\begin{align}
\rho \; \frac{d}{dt} \left[ \: PV(\psi) \: \right] & 
  \: = \; \frac{1}{{\rho}^{2}}\;  \boldsymbol{\nabla}(\psi) \: . \: 
          \left[ \: \boldsymbol{\nabla}(\rho)  \times \boldsymbol{\nabla}(p) \: \right]
   \; + \; \boldsymbol{\zeta}_a\: .\: \boldsymbol{\nabla}\!\left( \frac{d \psi}{dt} \right) 
   \; + \: \left( \boldsymbol{\nabla} \!\times\! \boldsymbol{F} \right) . \boldsymbol{\nabla}(\psi) 
  \: . \label{def_dPVdt_psi}
\end{align}

It is mentioned in H85 that ``{\it Normally in meteorology, $\psi$ is taken to be the potential temperature, although it can be equally well be taken to be the specific entropy or any other function of potential temperature \/}'.
The original Ertel formulation (1942, English translations in Schubert {\it et al.\/}, 2004) indeed  corresponds to the case $\psi=\theta$, where the dry-air potential temperature is equal to
\begin{align}
  \theta  & 
  \: = \; T \:\left(\frac{p_0}{p}\right)^{\kappa}
  \: . \label{def_Theta}
\end{align}
The $PV$ thinking popularized by H85 makes use of this formulation which demonstrates at the same time a Lagrangian conservation property and a principle of invertibility for adiabatic, reversible and frictionless motion of dry air.

The reason why the choice of $\psi=\theta$ annihilates the solenoidal term in the first line of the R.H.S. of  (\ref{def_dPVdt_psi}) is that $\theta$ can be expressed in terms of $\rho$ and $p$ only, according to the dry-air equation of state which may be written
\begin{align}
  p  & 
  \: = \; \rho \: R_d \: T
  \: = \; \rho \: R_d \: \theta \:\left(\frac{p}{p_0}\right)^{\kappa}
  \: , \label{def_dry_state} \\
  \theta (\rho, p)  & 
  \: = \; \frac{p}{\rho \: R_d}  \: \left(\frac{p_0}{p}\right)^{\kappa}
  \: . \label{def_dry_theta}
\end{align}
The second line in the R.H.S. of  (\ref{def_dPVdt_psi}) also cancels out for the adiabatic, reversible and frictionless motion of dry air, because the first term depends on the material derivative of $\theta$ and is  equal to zero since the dry-air entropy $s = c_{pd}\:\ln(\theta)+C^{ste}$  corresponds to the property $d\,\theta/dt = (\theta/c_{pd}) \: ds/dt =0$.
The last term of  (\ref{def_dPVdt_psi}) depends on the curl of the friction force and it vanishes for frictionless motion.
As a consequence, the material derivative (\ref{def_dPVdt_psi}) is  indeed equal to zero for $\psi = \theta$ and during the reversible, adiabatic and frictionless motion of a parcel of dry air.

Ertel's formulation is often used to study the thermodynamical properties of moist atmosphere, though it is based on $\theta$ and not on a moist counterpart.
It is assumed that the impact of water phase change can be represented by the diabatic term ${d \, \theta}/{dt}$ (i.e. condensation-evaporation and radiative  processes).
However, the corresponding inversion method does not generate tendencies for water species and the anomalies of $PV(\theta)$ are of small amplitude in the lower troposphere, creating difficulties for (or even an impossibility of) the analsis of the signals associated with frontal or convective moist systems.

In order to avoid these drawbacks, moist versions of the potential vorticity have been used in studies of atmospheric systems.
Various moist formulations of $PV$ have been built and tested by following the advice of H85, with $\psi$ replaced by the wet bulb ($\theta'_w$), moist entropic ($\theta_S$), equivalent ($\theta_e$), saturation equivalent ($\theta_{es}$) or virtual ($\theta_{v}$) potential temperatures.

Bennetts and Hoskins (1979, hereafter referred to as BH79) and Emanuel (1979, 1983) have investigated the  generation of conditional symmetric instability in baroclinic systems as a possible mechanism for the formation of frontal rain-bands.
It is suggested that negative moist potential vorticity computed with either $\psi=\theta'_w$, $\theta_e$ or $\theta_{es}$ is a sufficient condition for two-dimensional frictionless conditional symmetric instability.
However, none of these potential temperatures fulfills the demand to verify, at the same time, a moist conservative property and an invertibility principle.

More precisely, it is shown in S01 and Schubert (2004, hereafter referred to as S04) that, even if the equivalent potential temperatures $\theta_e$ or $\theta_{es}$ are approximately conserved during the motions of parcels of moist saturated air, and as far as the moist entropy may indeed be approximated by $c_{pd}\:\ln(\theta_e)$ or $c_{pd}\:\ln(\theta_{es})$, the solenoidal term in the first line of the R.H.S. of (\ref{def_dPVdt_psi}) does not cancels out.
This is an cumbersome source/sink term which cannot be associated with any conservative law.
Moreover, $PV(\theta_e)$ do not verify an invertibility principle.

In a symmetric way, it is shown in S01 and S04 that the choice of the virtual potential temperature $\psi=\theta_v$ annihilates the solenoidal term in (\ref{def_dPVdt_psi}) and that $PV(\theta_v)$ demonstrates an invertibility principle.
However, the corresponding diabatic term ${d \, \theta_v}/{dt}$ does not receive a clear physical interpretation, since it does not correspond to any moist conservative law.
In fact, $\theta_v$ is the key quantity used for buoyancy analyses, but it cannot be related to the moist entropy or to the Second Law, or to the conservation of matter for the mixing of dry air and total water species.

It is mentioned in Davies and Wernli (1997) that the $PV$ perspective is founded in three notions: \\
i) $PV$ conservation in the adiabatic and frictionless limit; \\
ii) $PV$ inversion in the balance-flow limit to obtain the associated flow and thermal pattern;\\
iii) $PV$ partition of the field into coherent distinctive elements of potential vorticity and potential temperature.
The aim of the present paper is to explore mainly the third notion.

The article is organized as follows.
Existing moist potential vorticities are described in Section~\ref{section_moist_PV}.
The main properties of the specific moist entropy, computed in M11 in terms of $\theta_s$, are described in Section~\ref{section_moist_S}.
The new moist potential vorticity $PV_s$ is defined in Section~\ref{section_PVs}, together with the virtual component $PV_v$ and with the justification for the use of a corresponding new potential vorticity unit (PVUS).
An approximate formulation for $PV_s$ is derived in Appendix~B, which uses the approximation of $\theta_s$ by $(\theta_s)_1$ described in M11.
The numerical experimentation is introduced in Section~\ref{section_numerical_model}, where some isobaric and iso-$PV$ charts are shown.
The cross-sections of potential temperatures are presented in Section~\ref{section_Cross_THETA}.
Isobaric charts and cross-sections of the new potential vorticity components are depicted in Section~\ref{section_PV_charts} and \ref{section_PV_cross}, respectively.
The possibility that the paradigm of slantwise convection may be revisited by the use of $PV_s$ and $\theta_s$ is analyzed in Section~\ref{section_slantwise}.
Charts of low-levels potential vorticity are shown in Section~\ref{section_PV_series} for a large northern midlatitudes Atlantic domain and for two successive forecasts.
Hints for next inversion methods based on the use of $PV_s$ are described in Section~\ref{section_inversion_method} and in Appendices~C and D, for both non-saturated and saturated moist air.
Use of new water component is suggested defined by $PV_q = PV_s - PV_v$ which highlight the negative lower-level $PV$ signals associated with moist processes.
It is explained that the magnitude of negative values and of solenoidal terms are smaller with the use of $\theta_s$ than with the use of $\theta_e$, two interesting features which could motivate a revival of interest in the search for a moist-air inversion tool applied directly to $PV_s$.
Finally, conclusions and outlines are presented in Section~\ref{section_conclude}.

%==========================================
\section{Existing moist potential vorticities.} % (Section 2)
%==========================================
\label{section_moist_PV}

The possibility of defining moist generalizations of Ertel's potential vorticity has been investigated by many authors.

The wet bulb potential temperature was used in BH79 to define the potential vorticity as
\begin{align}
q_w
 & = \: f \;\frac{g\:\rho}{\theta_0} \; PV(\,\theta'_w\,) 
  . \label{def_q_w}
\end{align}
In fact, $\theta'_w$ is conserved only during pseudo-adiabatic ascents of a saturated and precipitating convective parcel.
The conservative property verified by $\theta'_w$  corresponds to irreversible processes of open systems, during which the condensed water is continuously removed from the parcels, resulting in the non-conservation of the moist entropy (a part of which is withdrawn by the precipitations).

Rivas Soriano and Garc\'{\i}a D\'{\i}ez (1997) replaced  $\theta'_w$ by the  entropic potential temperature defined  in Hauf and H\"{o}ller (1987), in order to take into account the impact of ice water.
The resulting potential vorticity is written as
\begin{align}
q_g 
 & = \: f \;\frac{g\:\rho}{\theta_0} \; PV(\,\theta_{S}\,) 
 \: ,  \label{def_q_g}
\end{align}
where the specific moist entropy is defined by $q_d\:s^{\ast}_r + q_d\:c_p^{\ast}\: \ln(\theta_{S}/T_0)$.
The drawback of this formulation is that the terms $s^{\ast}_r$ and $c_p^{\ast}$ both depend on $r_t$, and thus vary with the total water mixing ratio.
The consequence is that the entropic potential temperature $\theta_{S}$ is not trully synonymous with the moist entropy, and $q_g$ cannot establish a general Second-Law conservative property (see M11).

In geostrophic coordinates, the moist potential vorticity is defined in Emanuel {\it et al.\/} (1987) with the saturation moist entropy $s_e$ expressed by $c_{pd}\: \ln(\theta_{es})$, where the equivalent potential vorticity is written in terms of this saturation equivalent potential temperature $\theta_{es}$, leading to
\begin{align}
q_{ge} 
 & = \: \frac{g\:\rho}{f \: \theta_e} \; PV(\,\theta_{es}\,) 
   \label{def_q_ge} \: , \\
\theta_{es} & \approx \theta \: \exp
          \left[
                - \: \frac{L_{vap}\:r_{sw}}{c_{pd}\:T}
          \right]
   \label{def_theta_es} \: .
\end{align}

The  moist generalizations $PV(\theta_{v})$ and $PV(\theta_e)$ are tested in S01, where the virtual potential temperature is defined by
\begin{align}
  \theta_{v}  & 
  \: = \; 
   \theta \: \left(\: 1  + \delta\: q_v  - q_l  - q_i \: \right)
  \: , \label{def_Theta_v1}
\end{align}
and the equivalent potential temperature by
\begin{align}
  \theta_e  & 
  \: = \; 
  T_0 \; \exp(s_e/c_{pd})
  \: . \label{def_Theta_E}
\end{align}
The equivalent entropy $s_e$ appearing in (\ref{def_Theta_E}) is determined according to Emanuel (1994).
It is expressed by two different formulas, depending on saturation or non-saturation conditions.

The moist-air equation of state is expressed in S01 in terms of the virtual potential temperature, leading to
\begin{align}
  p  & 
  \: = \; \rho \: R_d \: \theta_{v} \:\left(\frac{p}{p_0}\right)^{\kappa}
  \: , \label{def_moist_state} \\
  \theta_{v} (\rho, p)  & 
  \: = \; \frac{p}{\rho \: R_d}  \: \left(\frac{p_0}{p}\right)^{\kappa}
  \: . \label{def_moist_theta_vl}
\end{align}
The important property described in S01 is that $\theta_{v}$ given by (\ref{def_moist_theta_vl}) only depends on $\rho$ and $p$, as in (\ref{def_dry_theta}) for the dry-air formulation $\theta$.
As a consequence the solenoidal term cancels out in (\ref{def_dPVdt_psi})  for $\psi = \theta_{v}$.
The corresponding moist $PV$ is defined in S01 by
\begin{align}
PV(\,\theta_{v}\,)   
 & = \, 
       \frac{1}{\rho}
       \: \; \boldsymbol{\zeta}_a\: . \: 
      \boldsymbol{\nabla}\!
      \left( \, \theta_{v} \, \right)
  . \label{def_PV_theta_vl}
\end{align}

It is then demonstrated in S01 and S04 that $PV(\theta_{v})$ follows an invertibility principle, indicating that the virtual potential temperature may be a good candidate to build a moist version for the potential vorticity.
However, the main drawback of the choice of $\psi = \theta_{v}$ is that the material derivative of $\theta_{v}$ does not correspond to a conservative property.
Indeed, the gradient of $d\,\theta_{v}/dt$ cannot be linked to the change in moist entropy, or to the conservation of dry air and water species, for instance.
As a result, the second line in the R.H.S. of the PV equation (\ref{def_dPVdt_psi}) does not cancel out for $\psi = \theta_{v}$.

It is suggested in S01 and S04 that it is possible to overcome this drawback by defining the moist PV equation (\ref{def_dPVdt_psi}) with $\psi = \theta_e$ or $\theta_{es}$.
Indeed, the material derivative $d\,\theta_{e}/dt$ or $d\,\theta_{es}/dt$ could be equal to zero for two different reasons.

This might be true for adiabatic and reversible processes if $\theta_e$ or $\theta_{es}$ were valid measurements of the moist air entropy.
This assumption is made for instance in Emanuel {\it et al.\/} (1987), with a moist entropy implicitly defined by
\begin{align}
s & \: = \; q_d\: s_e \: = \: q_d\:c_{pd}\: \ln(\theta_e)
  \:+\: q_d\:C_1 \:+\:q_t\:C_2
   \label{def_s_E94} \: .
\end{align}
The two terms $C_1$ and $C_2$ represent additional terms which are either constant or depends on $q_t$, with $q_t$ also assumed to be a constant conservative quantity.

However, neither $\theta_e$ nor $\theta_{es}$ are valid measurements of the more general moist-air specific entropy, as shown in M11.
The main problem is that $q_d$ and $q_t$ are located outside the logarithm of $\theta_e$ in (\ref{def_s_E94}) and that a source of variability due to $q_d=1-q_t$ is thus missing in $\theta_e$.
Another impact of varying $q_d$ and $q_t$ in (\ref{def_s_E94}) is that $q_d\:C_1$ and $q_t\:C_2$ become variable terms which cannot be neglected, even if $C_1$ or $C_2$ were true constant terms.
Moreover, $c_{pd}$ is replaced by $c_{pd}+ r_t\: c_l$ in Emanuel (1994), with the impact of $r_t\: c_l$ also missing in $\theta_e$.

The facts that $\theta_e$ or $\theta_{es}$ are considered as conserved quantities, and that the material derivatives are almost equal to zero, correspond to pseudo-adiabatic conservative properties which are demonstrated by the potential temperature $\theta'_w$ in convective or frontal regions.
It is assumed that the liquid water droplets or ice crystals are withdrawn from cloud in the form of precipitations species as far as they are created by condensation processes.
These are irreversible kind of processes, and parts of the entropy are withdrawn by precipitation.

The consequence is that the conservative properties assumed for $\theta_e$ or $\theta_{es}$ do not correspond to the Second Law of thermodynamics applied to a specific parcel of moist air.
It corresponds to the change in entropy expressed per unit mass of dry air of a moist parcel, and so making it useless for applications to a barycentric vision and to the computation of material derivatives of moist air.
Another drawback linked to the choices of $PV(\theta_e)$ or $PV(\theta_{es})$ is that the associated solenoidal terms does not cancel out, and that these equivalent potential vorticities are not invertible.

Another ``saturation equivalent'' potential temperature $\theta^{\star}_e$ is defined in S04.
It is built in order to cancel the solenoidal term, and like $\theta_v$ it is invertible under balanced conditions, but it does not have the correct limit in the dry-air case -- i.e. $PV(\theta)$ -- making it ``useless for general applications'', according to the author.

Taking into consideration the advantages and drawbacks described in S01 and S04, the method followed in the next section will consist of modifying the existing moist air approaches by seeking a new specific (i.e. per unit of moist air) $\psi$ function to be put into (\ref{def_PV_psi}) and (\ref{def_dPVdt_psi}) and which could satisfy moist conservative properties (i.e. for which changes in material derivative are caused by pure diabatic processes).

%======================================================
\section{The moist entropy.} %  (Section 3)
%======================================================
\label{section_moist_S}

The aim of the present article is to express the formulation (\ref{def_PV_psi}) in terms of the specific moist air entropy, namely with $\psi = s$.
The moist air entropy is expressed in M11 in terms of a potential temperature  $\theta_{s}$, according to the relationships
\begin{align}
s & = \: s_{ref} \:  + \: c_{pd} \: \ln(\theta_{s})
  \: , \label{def_s_ln_theta_s} \\
({\theta}_{s})_1   
   & = 
        \: \theta \;
         \exp \left( - \:
                     \frac{L_v\:q_l + L_s\:q_i}{{c}_{pd}\:T}
                \right)
       \; \exp \left( {\Lambda}_r\:q_t \right)
  \label{def_theta_s1} \: , \\
{\theta}_{s}   
   & = \;  ({\theta}_{s})_1  \;
        \left( \frac{T}{T_r}\right)^{{\lambda} \:q_t}
        \left( \frac{p}{p_r}\right)^{-\kappa \:\delta \:q_t}
     \left(
      \frac{r_r}{r_v}
     \right)^{\gamma\:q_t}
     \;
      \frac{(1+\eta\:r_v)^{\:\kappa \: (1+\:\delta \:q_t)}}
           {(1+\eta\:r_r)^{\:\kappa \:\delta \:q_t}}
  \: . \label{def_THs}
\end{align}
The formulation (\ref{def_s_ln_theta_s}) provides a definition (\ref{def_THs}) for $\theta_{s}$ which is different from the definition obtained for $\theta_{e}$ from (\ref{def_Theta_E}).

The dimensionless  ${\Lambda}_r$-term is defined in M11 by
\vspace{-1.5mm}
\begin{align}
{\Lambda}_r & = \; \frac{ (s_{v})_r - (s_{d})_r }{ c_{pd}} \;  \approx \: 5.87
  \: . \label{def_Lambda}
\end{align}
This term ${\Lambda}_r$ depends on the difference between the reference values $(s_{v})_r$ and $(s_{d})_r$ for dry air and water vapour, respectively.
Therefore ${\Lambda}_r$ depends on the reference values  $T_r$, $p_r$ and $e_r$.
However,  it is explained in M11  that the other terms in the full formulation (\ref{def_THs}) rearrange so that ${\theta}_{s}$ is really independent on the reference values.
It is thus equivalent to the specific moist-air entropy state function and it may be computed locally.

A justification for studying ${\theta}_{s}$ as a relevant meteorological variable is given in M11.
Investigations of the thermodynamical internal structure of several  marine stratocumulus demonstrate that, even if the clear-air and in-cloud vertical profiles of the basic variables are not homogeneous, the vertical  gradients of thermal and water content properties combine so that the moist entropy and ${\theta}_{s}$ are almost constant within the PBL and the top-PBL entrainment region.

It is shown in M11 and in Marquet and Geleyn (2013) that ${\theta}_{s}$  can be accurately represented by the approximation $ ({\theta}_{s})_1$  given by (\ref{def_theta_s1}) and with the typical value of about $5.87$ for ${\Lambda}_r$.
If ice water content is zero, this approximate formulation only depends on the two well-known moist conservative variables $\theta_l$ and $q_t$ (Betts, 1973), with the result
\begin{align}
 ({\theta}_{s})_1
   & = \: \theta_l
       \; \exp \left( {\Lambda}_r\:q_t \right)
  \: . \label{def_theta_s1_Betts}
\end{align}

In order to avoid misinterpretations, the exact formula  (\ref{def_THs}) for  ${\theta}_{s}$  will be used in the next sections to define and compute the  new moist potential vorticity components $PV(s)$, denoted by $PV_s$.
The small impact of the approximation of ${\theta}_{s}$ by $({\theta}_{s})_1$  will be demonstrated by comparing the corresponding exact and approximate versions of $PV_s$ in Section~\ref{section_PV_cross} and in Appendix~B.

%======================================================
\section{The new moist $PV$ components.} %  (Section 4)
%======================================================
\label{section_PVs}

The aim of this section is to compare the formulation $PV(\theta_{v})$ studied in S01 with a new formulation $PV(s)$ computed with the moist entropy given by (\ref{def_s_ln_theta_s})-(\ref{def_Lambda}).

Since $c_{pd}$ and  $s_{ref}$  are constant terms, the quantity $PV(s)$ is equal to $c_{pd}\: PV[\, \ln(\theta_s)\,]$.
The moist entropy represented by $c_{pd}\: \ln(\theta_s)$ is thus a fundamental specific quantity associated with the Second Law.

Let us  consider the identity
\begin{align}
c_{pd} \: \ln(\theta_{s}) & = \; c_{pd} \: \ln(\theta_v) \: + \: c_{pd} \: \ln(\theta_{s}/\theta_v)
  \: . \label{def_ln_theta_vl_ln_theta_s}
\end{align}
The choices of $\psi = c_{pd} \: \ln(\theta_{v})$ and $\psi = c_{pd} \: \ln(\theta_{s})$ correspond to the exact differential properties
\begin{align}
c_{pd} \: d\,[\,\ln(\theta_{v})\,] & = c_{pd} \: \frac{d\,\theta_{v}}{\theta_{v}}
 \: , \label{def_d_ln_theta_vl} \\
c_{pd} \: d\,[\,\ln(\theta_{s})\,] & = c_{pd} \: \frac{d\,\theta_{s}}{\theta_{s}}
 \: . \label{def_d_ln_theta_s}
\end{align}
In contrast with the differential $d\,\theta_{v}$ and the formulation $PV(\theta_{v})$ studied in S01, the division by the potential temperature $\theta_{v}$ and the multiplication by $c_{pd}$ in (\ref{def_d_ln_theta_vl}) imply a change of unit and a change in magnitude.

More precisely, if $PV(\theta)$ and $PV(\theta_{v})$ are expressed with the standard unit of $10^{-6}$~m${}^{2}$~K~s${}^{-1}$~kg${}^{-1}$ (i.e. the PVU), a possible unit for $PV(s)$ may be  $10^{-6}$~m${}^{4}$~K${}^{-1}$~s${}^{-3}$~kg${}^{-1}$.
This generates a magnifying factor equal to $c_{pd}/\theta_{v} \approx 3.5$ to $3$ in the troposphere and the lower stratosphere, respectively.
This factor decreases toward the values $2.5$ and $2$ in the upper stratosphere.

Let us define the dry-air entropy, virtual and moist entropy potential vorticities by
\begin{align}
PV_{\theta} \: = \: \frac{c_{pd}}{3} \: PV[\,\ln(\theta)\,]  
 & = \, 
     \frac{c_{pd}}{3\:\theta} \:
      PV( \theta )
  \label{def_PV_ln_theta} \: , \\
PV_{v} \: = \: \frac{c_{pd}}{3} \: PV[\,\ln(\theta_{v})\,]  
 & = \, 
     \frac{c_{pd}}{3\:\theta_{v}}
     PV( \theta_{v} )
  \label{def_PV_ln_theta_vl} \: , \\
PV_{s} \: = \: \frac{c_{pd}}{3} \: PV[\,\ln(\theta_{s})\,]  
 & = \, 
     \frac{c_{pd}}{3\:\theta_{s}}
     PV( \theta_{s} )
  . \label{def_PV_ln_theta_s}
\end{align}
Division by the factor $3$ in the new definitions (\ref{def_PV_ln_theta})-(\ref{def_PV_ln_theta_s}) is convenient in practice, in order to recover standard values close to $1.5$ for $PV_{\theta}$ at the tropopause.
This corresponds to a modified moist entropy potential vorticity unit (PVUS) equal to $\:3 \, 10^{-6}$~m${}^{4}$~K${}^{-1}$~s${}^{-3}$~kg${}^{-1}$.

The component $PV_{s}$ depends, from (\ref{def_PV_ln_theta_s}), on the gradient of $\ln(\theta_{s}) = (s-s_{ref})/c_{pd}$, and thus on the gradient of the specific moist entropy, since $s_{ref}$ and $c_{pd}$ are two constants.
Therefore,  $PV_{s}$ observes the same conservative behaviour as the specific moist entropy.

According to S04 it is important to confirm that the dry-air limit of $\theta_s$, which corresponds to $q_v=q_l=q_i=0$, is equal to $\theta$.
This expected result is true, leading to $PV_{s} = PV_{v} = PV_{\theta}$ for a dry-air parcel, a quantity which is proportional to the expected Ertel's formulation $PV({\theta})$, but expressed in terms of the dry entropy and represented by (\ref{def_PV_ln_theta}).
The differences with the Ertel's formulation are the coefficient $c_{pd}/(3\:\theta)$ and the corresponding change of unit (PVU replaced by PVUS).

The formulation (\ref{def_PV_ln_theta_vl}) for $PV_{v}$ can be separated into the sum of two components.
The method is to take the logarithm of $\theta_{v}$ given by (\ref{def_Theta_v1}).
The result is the sum
\begin{align}
PV_{v}  & \: = \; 
 PV_{\theta}
 \; + \: 
 \frac{c_{pd}}{3} \; 
 \frac{PV\!\left( \delta\: q_v  -  q_l  - q_i \right)}
      {\left( 1  + \delta\: q_v  - q_l  - q_i \right)}
  \: . \label{def_PV_vl_bis}
\end{align}
The first component $PV_{\theta}$ is given by (\ref{def_PV_ln_theta}) and depends on the Ertel's potential vorticity formulation $PV(\theta)$.
The second component  in (\ref{def_PV_vl_bis}) depends on $PV[\: \delta\: q_v  - q_l  - q_i \:]$, a quantity which depends only on the gradient of water contents.

It is possible to compute numerical values of $PV_{s}$ given by (\ref{def_PV_ln_theta_s}) and with the full formulation (\ref{def_THs}) for $\theta_s$, as demonstrated in the numerical experimentations shown in the next sections.
It is more difficult to understand and interpret accurately all the physical properties of the components $PV_{s}$, because the expression of $\theta_s$ is really complicated, due to all terms in (\ref{def_THs}) which multiply $({\theta}_{s})_1$.
The component  $PV_{s}$  may be approximated, if needed, by assuming that the moist entropy potential temperature can be approximated by $ ({\theta}_{s})_1$ given by (\ref{def_theta_s1}), as shown in Appendix~B.

%======================================================
\section{Numerical experimentations.} %  (Section 5)
%======================================================
\label{section_numerical_model}

The French ARPEGE-IFS operational NWP model is used with the stretched and tilted pole configuration and with a zoom coefficient of $2.4$ over  France.
The equivalent truncation is close to T$2000$ with a maximum resolution of about $10$~km over France, Spain and the UK.
The resolution is better than $20$~km over the whole northern mid-latitudes Atlantic domain. 
The truncation is  close to T$333$ with a resolution of $60$~km over Australia.
The model is run with $70$ hybrid vertical levels and with a reduced Gauss grid of $800$ latitudes.

% ======================================
% ============ Figure 1 ================
% ======================================
\begin{figure}[hbt]
\centering 
\includegraphics[width=0.49\linewidth,angle=0,clip=true]{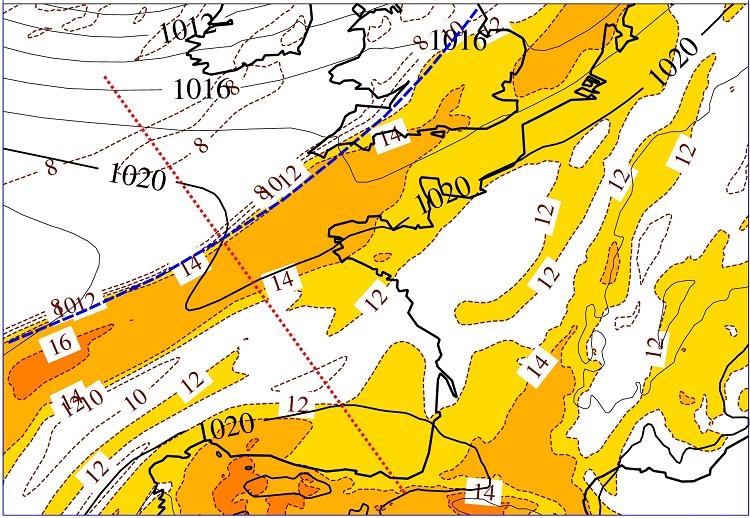}
\includegraphics[width=0.49\linewidth,angle=0,clip=true]{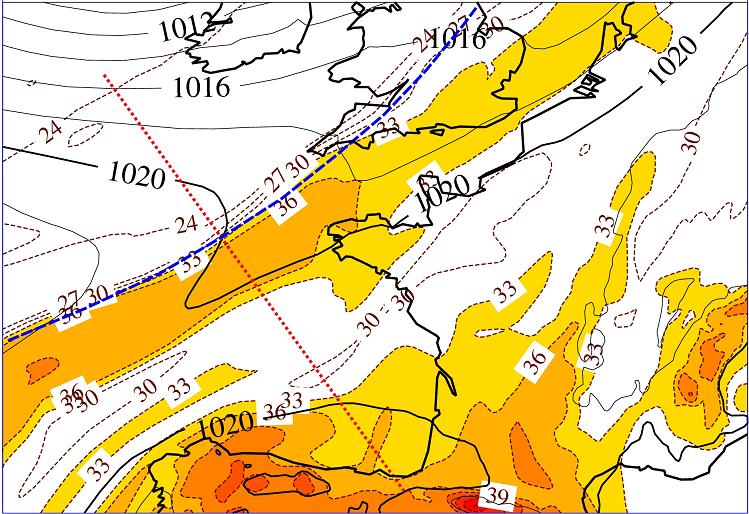}\\
(a) MSLP and $\theta'_w$ \hspace{3cm} (b) MSLP and $\theta_s$
\caption{\it \small
Operational outputs from the ARPEGE-IFS model for $3$~hours forecasts from $0000$~UTC on 21 September 2011.
The horizontal resolution of the post-processed grid is $0.1^{\circ}$ ($\approx 10$~km).
Mean sea level pressure (contours every 2~hPa, bold contour for $1020$~hPa), and potential temperatures (a) $\theta'_w$ at $850$~hPa (contour every $2^{\circ}$C, shaded above $12^{\circ}$C), and (b) $\theta_s$ at $850$~hPa (contour every $3^{\circ}$C, shaded above $33^{\circ}$C).
The dotted straight line represents a vertical cross-section extending from the west of Ireland ($49^{\circ}$N - $10^{\circ}$W) to northeast of Spain ($46^{\circ}$N - $6^{\circ}$W).
The dashed line represents the $850$~hPa northern frontal limit.
\label{Fig_MSLP_THPW850}}
\end{figure}

The mean sea level pressure and the $850$~hPa $\theta'_w$ fields are depicted in Figure~\ref{Fig_MSLP_THPW850}(a) for the 21st of September, 2011 ($3$~hour forecast from $0$~UTC).
The frontal region extends from the southwest to the northeast of the domain, where it is located over the Channel and the North Sea.
The frontal region is roughly delimited by the shaded values of $\theta'_w$  higher than $12^{\circ}$C.
The northern frontal limit is outlined at $850$~hPa as a dashed line, whereas the surface front is associated with the  mean sea level pressure trough located in the middle of the $850$~hPa frontal region.

The dotted straight line  extending from the west of Ireland to the north of Spain represents the cross-section which will be studied in the next sections.
It crosses the frontal region at $1020$~hPa to the west of  France and south of Ireland.

The $850$~hPa $\theta_s$ field is depicted in Figure~\ref{Fig_MSLP_THPW850}(b).
It is similar to the $\theta'_w$ field, provided that the frontal region is defined by values of $\theta_s$ higher than $33^{\circ}$C (versus above $12^{\circ}$C for $\theta'_w$) and if contours of $\theta_s$ are  plotted every $3^{\circ}$C (versus every $2^{\circ}$C for $\theta'_w$).

% ======================================
% ============ Figure 2 ================
% ======================================
\begin{figure}[hbt]
\centering
\includegraphics[width=0.65\linewidth,angle=0,clip=true]{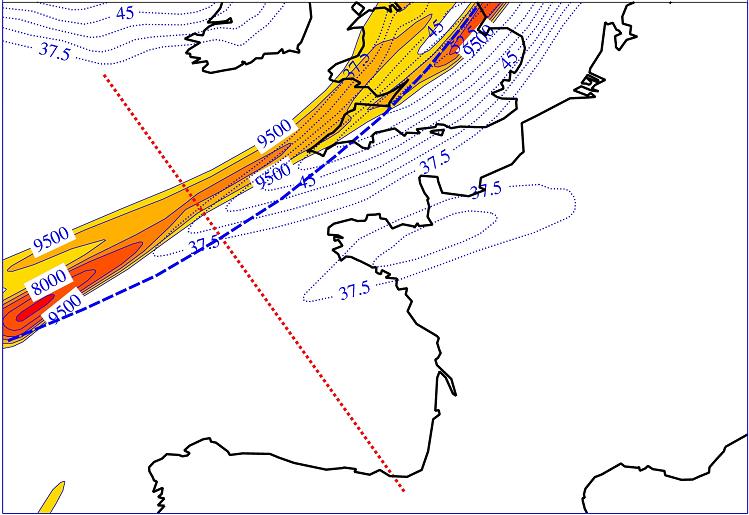}
\caption{\it \small
Area as Figure~\ref{Fig_MSLP_THPW850}, but showing the $250$~hPa wind speed (dotted contours every $2.5$~m~s${}^{-1}$ above $37.5$~m~s${}^{-1}$), and the height of the $1.5$~PVU level, computed with the usual Ertel's version $PV(\theta)$ (shaded areas from  $5000$ to $8000$ every $1000$~m, then every $500$~m up to $9500$m).
\label{Fig_V250_Z15PVU}}
\end{figure}

The upper-levels dynamical fields are shown in  Figure~\ref{Fig_V250_Z15PVU}.
The patterns of the geopotential on the $PV(\theta)=1.5$~PVU surface and the 250 hPa wind speed exhibit elongated features of low values for the $1.5$~PVU geopotential  and a significant jet area, both aligned near and to the north of the  $850$~hPa front.
These features indicate that the cross-section is located accross the upper air anomalies and to the southwest (rear) of the Jet.

%======================================================
\section{Cross-sections of potential temperature.} %  (Section 6)
%======================================================
\label{section_Cross_THETA}

\begin{figure}[hbt]
\centering
\includegraphics[width=0.49\linewidth,angle=0,clip=true]{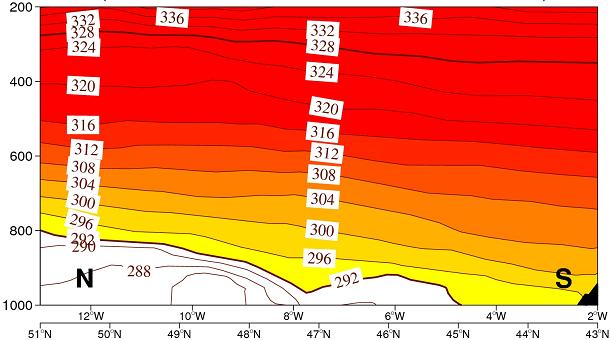}
\includegraphics[width=0.49\linewidth,angle=0,clip=true]{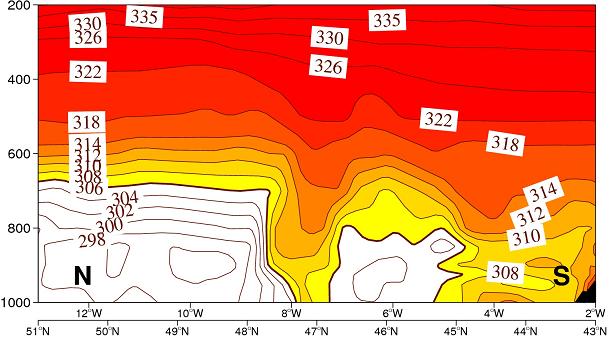}
\\ (a) $\theta_v$  \hspace{6cm}  (b) $\theta_s$ \\
\includegraphics[width=0.49\linewidth,angle=0,clip=true]{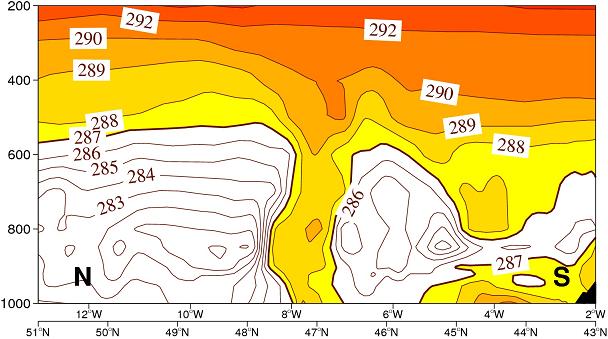}
\includegraphics[width=0.49\linewidth,angle=0,clip=true]{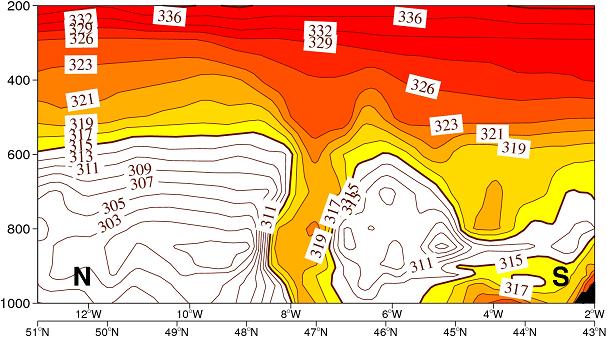}
\\ (c) $\theta'_w$   \hspace{6cm} (d) $\theta_e$
\caption{\it \small
Vertical cross-section computed along the dashed lines depicted in Figures~\ref{Fig_MSLP_THPW850} and \ref{Fig_V250_Z15PVU} for (a) $\theta_v$, (b) $\theta_s$, (c) $\theta'_w$ and (d) $\theta_e$ (Bolton, 1980).
The orography of Spain appears at the Southern end.
Potential temperatures are in K.
Frontal regions are delimited by the shaded regions at $\theta_s > 33^{\circ}$C, $\theta'_w > 14^{\circ}$C and $\theta_e > 42^{\circ}$C.
Post-processed levels are at $1000$, $950$, $925$, $900$, $850$, then every $100$~hPa from $800$ to $200$~hPa. 
\label{Fig_coupes_Theta}}
\end{figure}

Cross-sections are plotted in Figures~\ref{Fig_coupes_Theta}(a)-(d) for the potential temperatures $\theta_v$,  $\theta_s$, $\theta'_w$ and $\theta_e$, respectively.
Numerical values of $\theta_e$ are computed according to Bolton (1980).
The location of the vertical cross-section is depicted in Figures~\ref{Fig_MSLP_THPW850} and \ref{Fig_V250_Z15PVU}.

The patterns of the frontal region are described differently by each potential temperature.
There is almost no signal in (a) with the virtual potential temperature $\theta_v$.
This confirms that the difference between $\theta$ and $\theta_v$ and the impact of moist air on $\theta_v$ must be small.

% ======================================
% ============ Figure 3 ================
% ======================================

The cross-sections of $\theta'_w$ and $\theta_e$ exhibit almost the same general pattern in (c) and (d), except that the contours and shaded regions correspond to different values.
The frontal limits may be associated with the bold lines $\theta'_w = 14^{\circ}$C$\:\approx 287$~K  or $\theta_e=42^{\circ}$C$\:\approx 315$~K, with $850$~hPa fronts located in the middle of the cross-section, on each sides of the longitude $8$~W.

Almost the same frontal region can be seen in (b) for $\theta_s$ and below the level $800$~hPa. 
The bold lines $\theta_s=33^{\circ}$C$\:\approx 306$~K delimit almost the same fronts as the bold lines of the pseudo-adiabatic potential temperatures ($\theta'_w$ and $\theta_e$), although the fronts are less vertical with $\theta_s$ than for the pseudo-adiabatic potential temperatures.

The vertical gradients of $\theta_s$ are smaller than the one observed for $\theta'_w$ and $\theta_e$ outside the frontal region and in the lower part of the planetary boundary layer (PBL; below $850$~hPa).
Moreover the marked minimum observed close to the level  $850$~hPa for both $\theta'_w$ and $\theta_e$ almost disappears for $\theta_s$.
This corresponds to the well-mixed regime already observed for the moist-air entropy in the PBL and at the top of te PBL of marine stratocumulus (M11), but verified here for the cold sectors located north and south to the frontal region, where moist turbulence and shallow convection are both active.

The differences are more important  above $700$~hPa or within the frontal region.
The pseudo-adiabatic potential temperatures $\theta'_w$ and $\theta_e$ are almost conserved up to $600$~hPa within the frontal region, whereas a vertical gradient of $\theta_s$ exist over the the whole frontal region, from $900$~hPa to $400$~hPa.
These differences can be explained by the different impacts of physical processes.
On the one hand, specific moist entropy increases with height, because vertical gradients of $\theta_s$ demonstrate the impact of irreversible or pseudo-adiabatic processes on moist entropy.
On the other hand, $\theta'_w$ and $\theta_e$ are conserved by deep-convection saturated and precipitating processes, because they have been defined for this purpose as conserved quantities, but expressed per unit mass of dry air.

It is worth remarking that contours of  $\theta_s$ in (b) are different from those of $\theta_v$  in (a) up to the $400$~hPa level, and especially in the moist warm sector.
This can be explained by the term $\exp({\Lambda}_r \: q_t)$ which multiplies $\theta_l$ -- i.e. almost  $\theta$ -- and  impacts largely on $\theta_s$ in spite of the decreasing values of $q_t$, due to the large value of ${\Lambda}_r \approx 5.87$ which is about $10$ times larger than $\delta \approx 0.608$, i.e. the term acting on $q_v$ in the moist formulation (\ref{def_Theta_v1}) for $\theta_v$.
The same impacts are observed in the warm sector of $\theta'_w$ and $\theta_e$ between $600$ and $400$~hPa.

%======================================================
\section{Isobaric charts  of potential vorticities.} %  (Section 7)
%======================================================
\label{section_PV_charts}

\begin{figure}[hbt]
\centering
\includegraphics[width=0.49\linewidth,angle=0,clip=true]{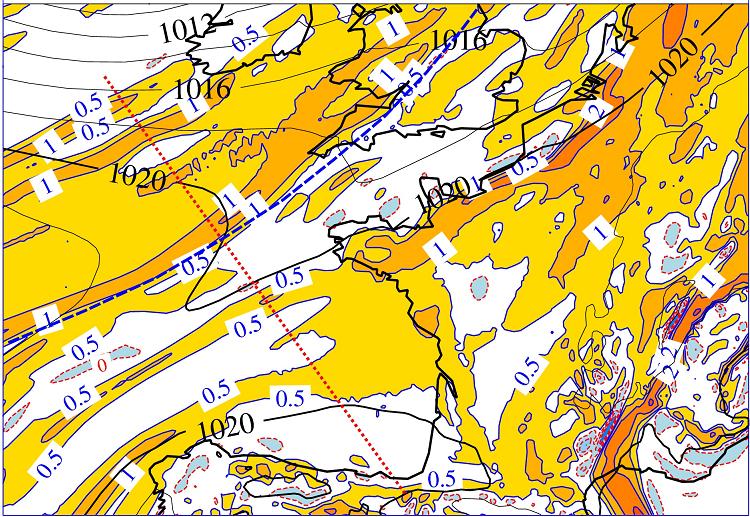}
\includegraphics[width=0.49\linewidth,angle=0,clip=true]{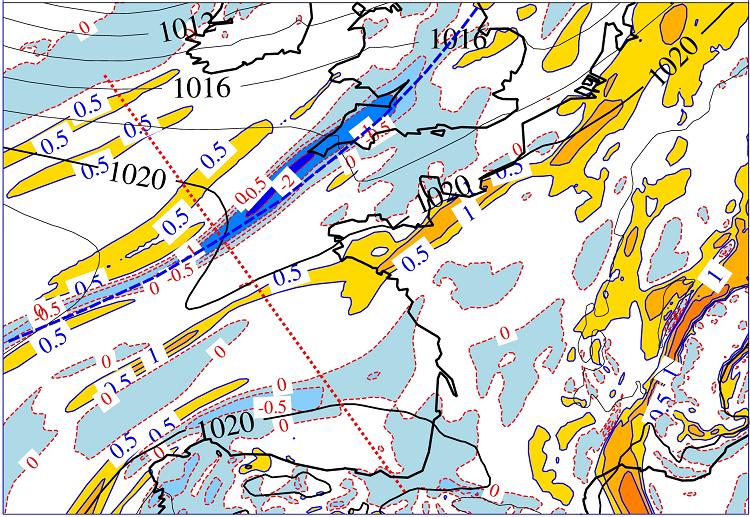}
\\ (a) MSLP and $PV_v$ \hspace{3cm} (b) MSLP and $PV_s$  \\
\includegraphics[width=0.49\linewidth,angle=0,clip=true]{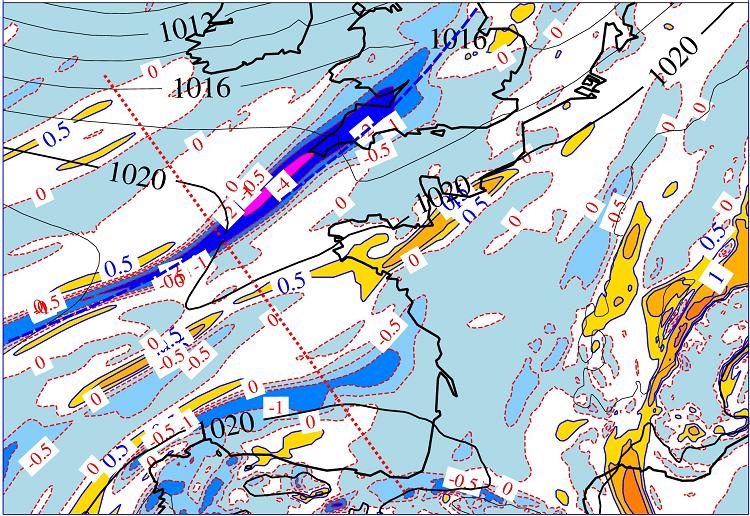}
\\ (c) MSLP and $PV_e$
\caption{\it \small
As in  Figure~\ref{Fig_MSLP_THPW850}, but showing the potential vorticity components plotted every $0.5$~PVUS at $850$~hPa for 
(a) the virtual component $PV_v$,
(b) the moist entropy component $PV_{s}$, and
(c) the equivalent component $PV_{e}$.
Shaded regions represent values of $PV>+0.5$~PVUS (solid contours) or $PV<0$~PVUS (dashed contours).
\label{Fig_cartes_PV_850}}
\end{figure}

The horizontal structures of the moist potential vorticities $PV_v$, $PV_e$ and $PV_e$  at $850$~hPa are shown in  Figure~\ref{Fig_cartes_PV_850}(a)-(c).

% ======================================
% ============ Figure 4 ================
% ======================================

In (a), the $PV_v$ field exhibits smooth structures, with shaded positive regions and values greater than $+0.5$~PVUS almost everywhere (except in the south of France).
The frontal region seems to be associated with an elongated region of positive values above $1$~PVUS located at the north of the northern limit (shaded line), with a less organized region located over the north-west of France.

In (b), the pattern of the $PV_s$ field is dominated by the white (unshaded) areas of moderate positive values (between $0$ and $+0.5$~PVUS).
An interesting feature is the marked minimum close to the frontal limit, with an elongated feature of  negative values less than $-2$~PVUS in the southwest of Cornwall.
This region is the one located below the entrance of the jet and almost below the elongated upper-air anomalies of $PV$.
The cross-section intersects the front within this elongated region of negative values of $PV_s$.
It may corresponds to slantwise convection instabilities, depending on the sign of the vertical gradient of $\theta_s$.
This possibility will be analyzed in more detail in Section~\ref{section_slantwise}.

In (c), the use of $\theta_e$ leads to large shaded regions of negative values of $PV_e$ (surrounded by dashed lines) almost everywhere.
There are only few positive shaded areas of $PV_e>+0.5$~PVUS (surrounded by solid lines).
The values of $PV_e$ are thus more negative than those for $PV_s$.
They are lower than $-4$~PVUS in the same elongated region already observed for $PV_s$, close to the frontal limit and below the upper-level anomalies.

The interesting feature is that values of $PV_s$ exhibit the same significant negative regions than those of $PV_e$, but with more moderate negative values, and with almost no negative values elsewhere in the domain.
This may be an important property to avoid the instabilities observed in inversion algorithms based on $PV(\theta_e)$, because too negative values make the inversion operator hyperbolic and unstable.

%======================================================
\section{Cross-sections of potential vorticities.} %  (Section 8)
%======================================================
\label{section_PV_cross}

% ======================================
% ============ Figure 5 ================
% ======================================
\begin{figure}[hbt]
\centering
\includegraphics[width=0.49\linewidth,angle=0,clip=true]{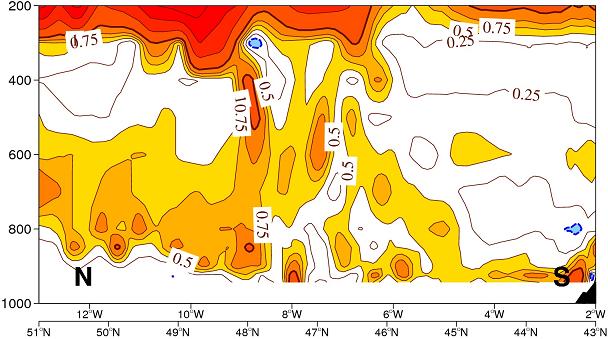}
\includegraphics[width=0.49\linewidth,angle=0,clip=true]{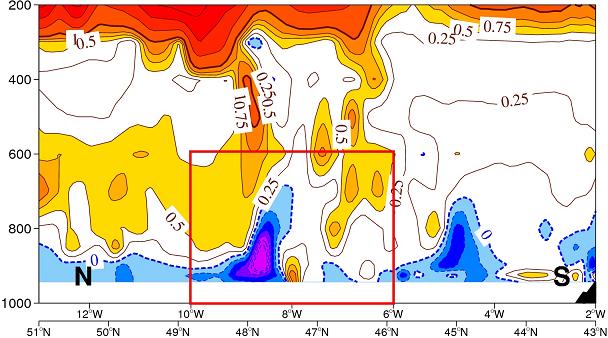}
\\ (a) $PV_v$   \hspace{6cm}  (b) $PV_s$  \\
\includegraphics[width=0.49\linewidth,angle=0,clip=true]{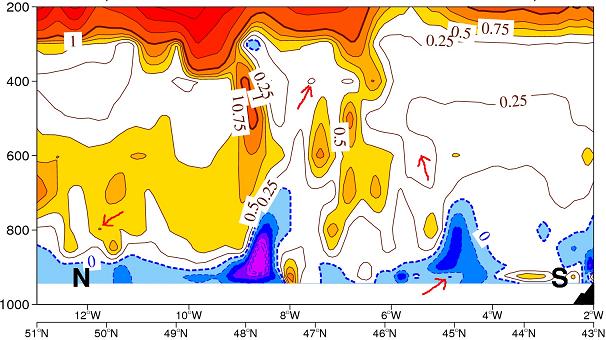}
\includegraphics[width=0.49\linewidth,angle=0,clip=true]{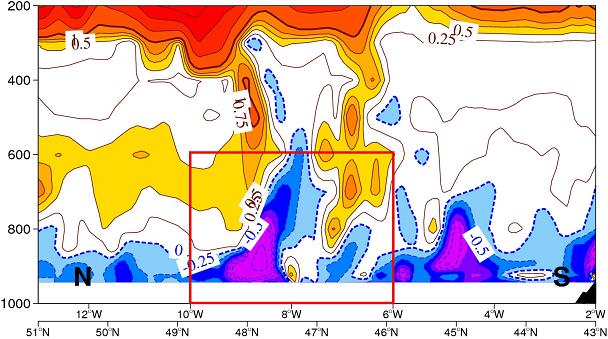}
\\ (c) $PV_{s1}$   \hspace{6cm}  (d) $PV_e$
\caption{\it \small
Cross-sections as in Figure~\ref{Fig_coupes_Theta}, but for the new potential vorticity components
(a) the virtual component $PV_v$,
(b) the moist entropy component $PV_{s}$,
(c) the approximate moist entropy component $PV_{s1}$,
(d) the equivalent component $PV_{e}$.
Potential vorticities are plotted at $0.25$~PVUS intervals between $-1.5$ and $1$~PVUS, with selected uneven values otherwise ($-10$, $-7.5$, $-5$, $-3$ and $1.5$, $3$, $5$, $7.5$, $10$ PVUS).
Values $>+0.5$~PVUS and $<0$~PVUS are shaded, with dashed contours $\leq0$.
The boxes in (b) and (d) represent a sub-region which is used in Figure~\ref{Fig_coupe_PV_dsdp}.
\label{Fig_coupes_PV}}
\end{figure}

% ======================================
% ============ Figure 6 ================
% ======================================
\begin{figure}[hbt]
\centering
\includegraphics[width=0.75\linewidth,angle=0,clip=true]{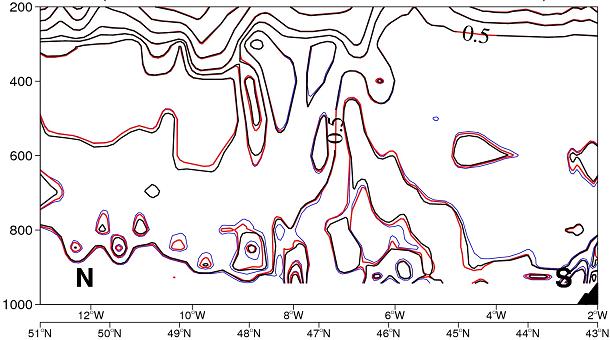}
\caption{\it \small
Cross-sections as in Figure~\ref{Fig_coupes_PV}(a), comparing Ertel's formulation $PV(\theta)$ (plotted every $0.25$~PVU), $PV_{\theta}$ and $PV_v$ (plotted every $0.25$~PVUS).
\label{Fig_coupe_PV_theta}}
\end{figure}

Vertical cross-sections as in Figure~\ref{Fig_coupes_Theta} are shown in  Figures~\ref{Fig_coupes_PV}(a)-(d)  for the moist potential vorticity functions $PV_v$, $PV_s$,  $PV_{s1}$ and $PV_e$, respectively.

The four moist potential vorticities  generate almost the same features above the mid-troposphere, at the tropopause and in the stratosphere, i.e. at and above $500$~hPa.
For instance, the stratospheric dry-air intrusions into the upper troposphere are almost the same for each potential vorticity (dark shading of values greater than $0.75$ and $1$~PVUS).

This cannot be explained by small values of $q_t$ observed above $500$~hPa, because they are large enough in the warm sector (located close to $8^{\circ}$W) to make $\theta_v$ different from the other components at $400$~hPa, as explained in section~\ref{section_Cross_THETA}.
The explanation is that the differences in potential temperatures mentioned in Section~\ref{section_Cross_THETA} are observed for the horizontal gradients, whereas the vertical gradients are almost the same.
And since vertical gradients of potential temperature multiply the large and dominating vertical component of absolute vorticity, the horizontal gradients of potential temperatures generate small terms in potential vorticity components.

Differences can be observed for moist-air potential vorticity components below $500$~hPa.
The wide  shaded regions observed in (a) corresponds to positive values of $PV_v$ greater than $0.5$~PVUS.
They are not observed for the other components in (b) to (d), where white areas correspond to values between $0$ and $0.5$~PVUS.

Larger differences are observed in the lower troposphere, below $700$~hPa.
Whereas values of $PV_v$ are positive everywhere, the frontal limits are associated with moderate negative values of $PV_s$ and with large negative values of $PV_e$.
This confirm the results observed in the $850$~hPa charts described in Section~\ref{section_PV_charts}.
These results are valid for other vertical levels, with a vertical extent of negative values of $PV_e$ which is larger than for the moist entropy component $PV_s$, especially within the warm sector, close to $8^{\circ}$~W.

The component $PV_{s1}$ depicted in (c)  is the same as $PV_s$ depicted in (b) and given by  (\ref{def_PV_ln_theta_s}), but with $\theta_s$ replaced by $(\theta_{s})_1$, giving $PV_{s1} = [\:{c_{pd}}/\{ 3\:(\theta_{s})_1 \}\:]\:PV[\: (\theta_{s})_1 ]$.
The comparison of the two exact and approximate versions shows very small differences.
Examples of these small differences are indicated by small arrows in (c).
This is a confirmation that the results obtained in M11, that $(\theta_{s})_1$ is indeed a relevant approximation for $\theta_s$, is also valid for the computations of the associated potential vorticities.

It is worth comparing the dry-air versions $PV(\theta)$ and $PV_{\theta}$ with the virtual component $PV_v$, in order to  demonstrate that the previous comparisons of $PV_s$ and $PV_e$ with $PV_v$ can be extended to $PV(\theta)$, the form of $PV$  used in most operational inversion tools.
The three fields of $PV(\theta)$, $PV_{\theta}$ and $PV_v$ are plotted  in Figure~\ref{Fig_coupe_PV_theta}.
The three fields are almost superimposed for the  selected contours ($0.5$, $1$, $1.5$, $3$, ...) and the lower-level contours of $0.5$~PVUS are almost the same.
This is another way to show that the impact of $q_t$ on $\theta_s$, and thus on $PV_s$, is much smaller than the impact of $q_t$ on $\theta_v$, and thus on $PV_v$, as explained at the end of Section~\ref{section_Cross_THETA}.

%======================================================
\section{Slantwise convection.} %  (Section 9)
%======================================================
\label{section_slantwise}

According to the criteria described in BH79, conditional symmetric instability and slantwise convection may occur in those regions where negative values of $PV$ correspond to positive vertical gradients of potential temperature.
In order to better motivate, and to find justification for, the use of $PV_s$ rather than $PV_v$ or $PV_e$ in future inversion tools, we plot on the same figures the potential vorticity components and the vertical gradient of the associated potential temperature.
The aim is to find areas where the criteria for slantwise convection might be observed for $\theta_s$, whereas they are not verified with $\theta$ or $\theta_e$.

% ======================================
% ============ Figure 7 ================
% ======================================
\begin{figure}[hbt]
\centering
\includegraphics[width=0.49\linewidth,angle=0,clip=true]{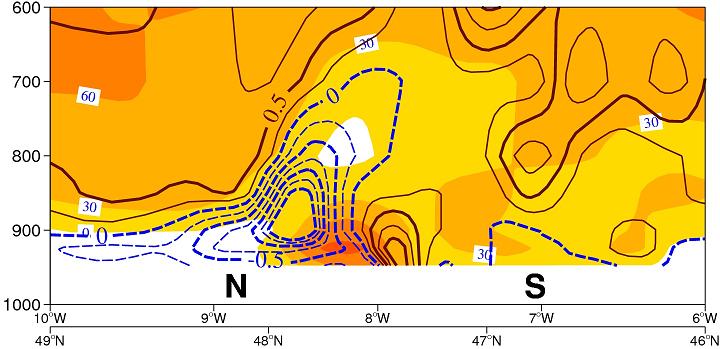}
\includegraphics[width=0.49\linewidth,angle=0,clip=true]{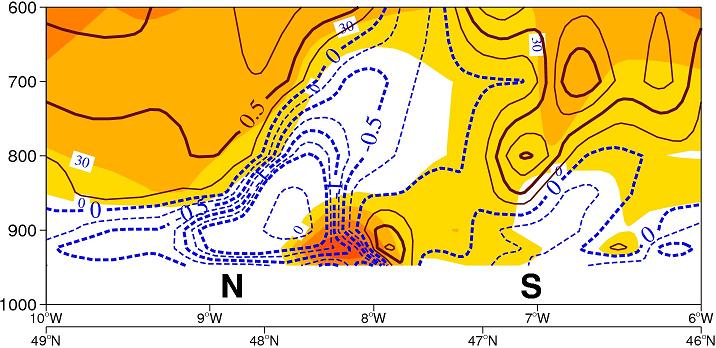}
\\ (a) $PV_s$  and $-\partial \theta_s/\partial p$ \hspace{3cm} (b) $PV_e$ and $-\partial \theta_e/\partial p$  \\
\includegraphics[width=0.49\linewidth,angle=0,clip=true]{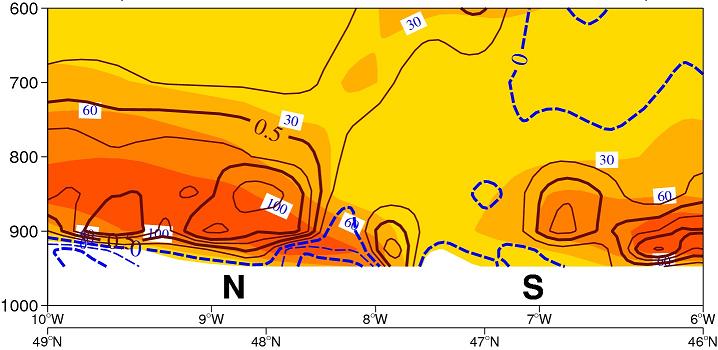}
\\ (c) $PV_{es}$ and $-\partial \theta_{es}/\partial p$  
\caption{\it \small
Cross-sections over the sub-area depicted by boxes in Figures~\ref{Fig_coupes_PV}(b) and (c) showing $PV$ components and 
vertical gradients ($-\partial \theta / \partial z$) for:
(a) $PV_s$ and $(\theta_s)$, 
(b) $PV_e$ and $(\theta_e)$ (Bolton, 1980), and
(c) $PV_{es}$ and $\theta_{es}$ given by (\ref{def_theta_es}).
Potential vorticities are plotted every $0.25$ PVUS intervals (solid contours if $>0$, dashed if $\leq 0$).
Vertical gradients are depicted as light shaded areas for positive values, and darker shading above $30$, $60$ and $100$ units of K~($1000$~hPa)${}^{-1}$.
\label{Fig_coupe_PV_dsdp}}
\end{figure}

$PV_s$ and the vertical gradient of $\theta_s$ are both plotted on  Figure~\ref{Fig_coupe_PV_dsdp}(a).
$PV_e$ and the vertical gradient of $\theta_e$ are plotted on Figure~\ref{Fig_coupe_PV_dsdp}(b).
These cross-sections correspond to the boxes depicted  in  Figures~\ref{Fig_coupes_PV}(b,d), between $10$ and $6^{\circ}$W and below $600$~hPa.

In (b), the negative values of $PV_e$ (dashed contours) correspond to negative vertical gradients of $\theta_e$ (white regions), indicating that vertical instabilities took place, rather than slantwise convection.
Conversely, in (a) negative values of $PV_s$ partly correspond to positive vertical gradients of $\theta_s$ (light shaded region).
This is true for the north and south frontal limits, at $9$ and $7^{\circ}$W.
This may indicates that conditional symmetric instability may occur and that slantwise convection may take place in these regions.

An interpretation for these results is that $\theta_s$ surfaces are more vertical than $\theta$ surfaces in the lower troposphere, but they are less vertical than $\theta'_w$ or $\theta_e$ surfaces, in a possible proportion of about $2/3$  between $\theta$ and the pseudo-adiabatic potential temperatures.
This proportion of $2/3$ is the one observed between ($\theta_l$, $\theta_s$, $\theta_e$) in M11 for the marine stratocumulus.

This property can be used to place the surface of constant $\theta_s$ on Figure~2 of BH79.
The $\theta_s$ surface must be closer to the absolute vorticity vector surface than the $\theta'_w$  surface.
This could explain why the negative values of  $PV_s$ are smaller than the one of $PV_e$, because this implies that the 3D absolute vorticity vector and the vector of 3D gradient of $\theta_s$ are more normal to each other than for the 3D gradient of $\theta'_w$ and $\theta_e$.

Another phenomenon may balance the impact due to change of slope of these surfaces.
A change in numerical value of the potential temperature may modify the magnitude of the gradient vector, and thus the value of potential vorticity.
It could be inferred from Figures~\ref{Fig_coupes_Theta}(b)-(d) that the magnitude of the gradient of $\theta_s$ is not larger than those of $\theta_e$, for instance at $48\:$N and below $900$~hPa, within that region of  negative values of $PV_s$ and $PV_e$.

The interpretation of the change of sign for the vertical gradients of $\theta_s$ and $\theta_e$ (close to $48^{\circ}$N and between $900$ and $800$~hPa) requires refined analyses and comparisons of the Figures~(\ref{Fig_coupes_Theta})(b) and (d).
The $\theta_e$ surfaces are oriented so that $\theta_e$ decreases with height, whereas $\theta_s$ surfaces are almost vertical,  so that $\theta_s$ is indeed almost neutral, or slightly increases with height.

The main problem to be solved is that the criteria for conditional symmetric instability and the possibility of slantwise convection cannot depend on the choice of $\theta_s$, $\theta'_w$ or $\theta_e$ to be used in the definitions of the associated vertical gradients and the potential vorticities.

A first remark is that the diabatic sources and sinks appearing in the entropy equation $d s/d t$ are better represented by $d \theta_s/d t$  than by $d \theta'_w/d t$ or $d \theta_e/d t$.
The explanation is given  in section~\ref{section_moist_PV}:   $\theta'_w$ or $\theta_e$ are expressed per unit mass of dry air, whereas  material derivatives and barycentric formulations require a specific definition for the moist entropy of moist air.
This requirement is displayed only by $\theta_s$ introduced in M11 and covered in Section~\ref{section_moist_S}.

The patterns of $PV_{es} = [\:c_{pd}/(3\:\theta_{es})\:]\:PV(\theta_{es})$ in Figure~\ref{Fig_coupe_PV_dsdp}(c), where $\theta_{es}$ is defined by (\ref{def_theta_es}),  are very different from those of $PV_e$ in (b).
Values of potential vorticities and vertical gradients are thoroughly modified in the lower troposphere.
The consequence is that $PV_{es}$ defined in terms of $\theta_{es}$ cannot be used from a forecaster's viewpoint.

It is thus tempting and logical to modify the method initiated in BH79 and to define a reduced static stability expressed by $N^2_s = (g/\theta_r)\:(\partial \theta_s/\partial z)$ and a corresponding entropic potential vorticity by $PV(\theta_s)$ or $PV_s$, with $\theta_s$ replacing $\theta'_w$, $\theta_e$ or $\theta_{es}$.
The  motivations are the same as in previous articles: to get a moist-air generalization of Ertel's potential vorticity by specifying the specific moist-air entropy, with the new result that $\theta_s$ should be used if all the properties of moist air entropy are to be taken into account.
It may be worthwhile to notice that $N^2_s$ somehow corresponds to the entropy part of the squared Brunt-V\"{a}is\"{a}l\"{a} frequency computed in Marquet and Geleyn (2013), provided that a positive factor in front of  $N^2_s$ is approximated by $1$, in the same way as the positive term $\Gamma_m/\Gamma_d$ in often discarded in the formulations which use $\theta'_w$ or $\theta_e$.

%======================================================
\section{Charts of low-levels potential vorticities.} %  (Section 10)
%======================================================
\label{section_PV_series}

\begin{figure*}[hbt]
\centering
\includegraphics[width=0.48\linewidth,angle=0,clip=true]{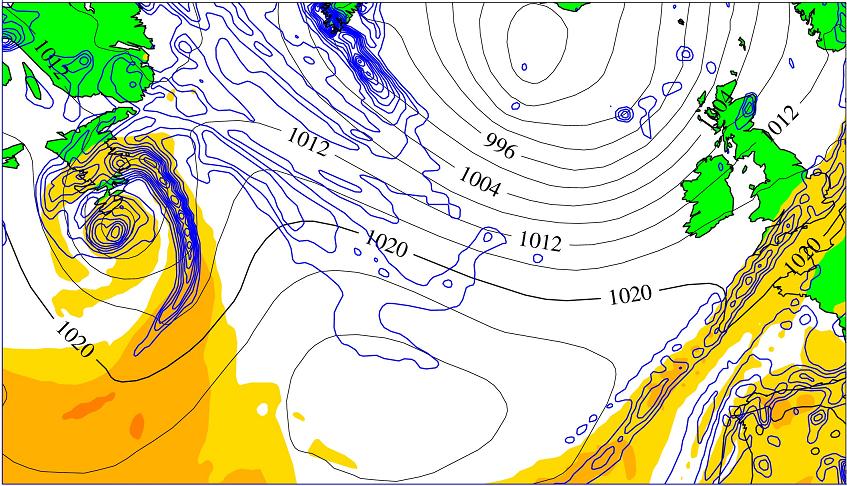}
\includegraphics[width=0.48\linewidth,angle=0,clip=true]{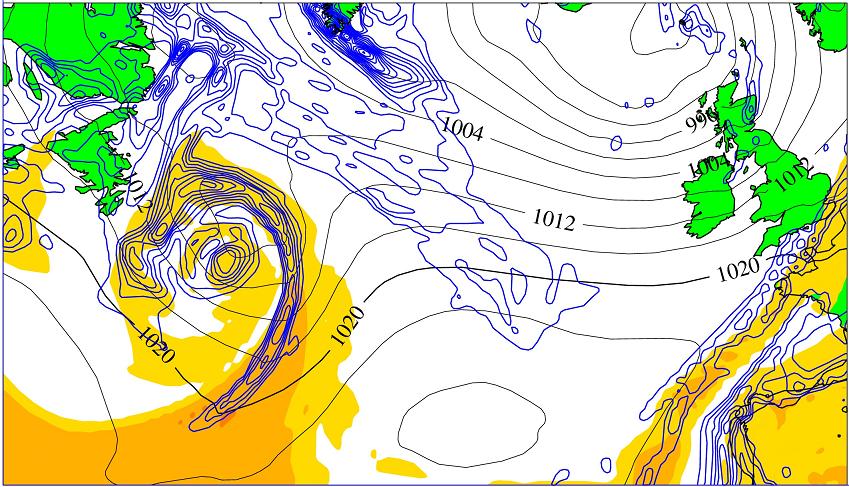}
\\ (a)  $PV_v$ / $0$~UTC \hspace{6cm} (b) $PV_v$ / $12$~UTC \\
\includegraphics[width=0.48\linewidth,angle=0,clip=true]{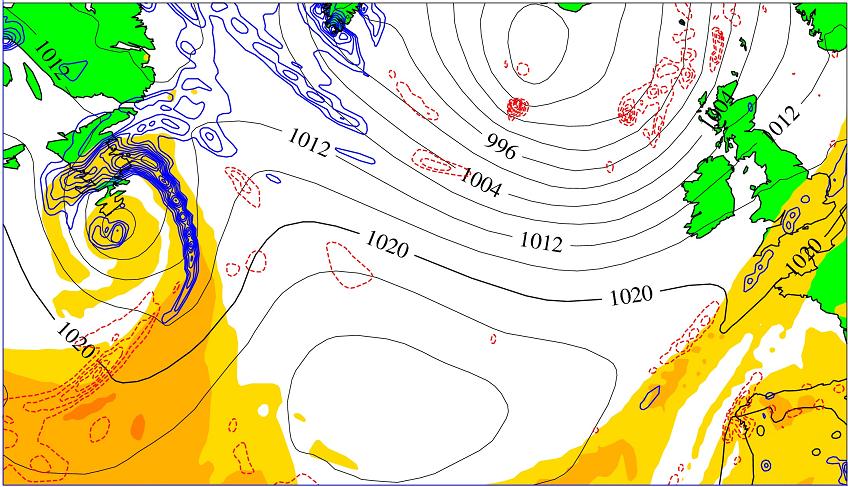}
\includegraphics[width=0.48\linewidth,angle=0,clip=true]{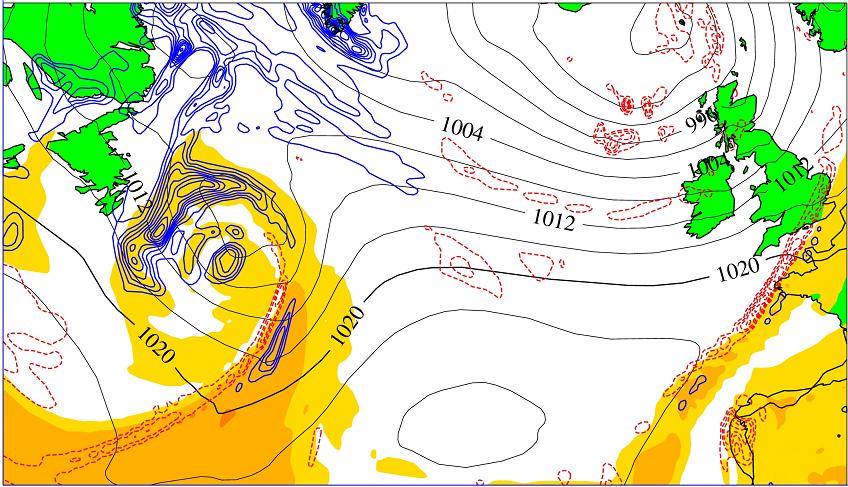}
\\ (c) $PV_s$ / $0$~UTC \hspace{6cm} (d) $PV_s$ / $12$~UTC \\
\includegraphics[width=0.48\linewidth,angle=0,clip=true]{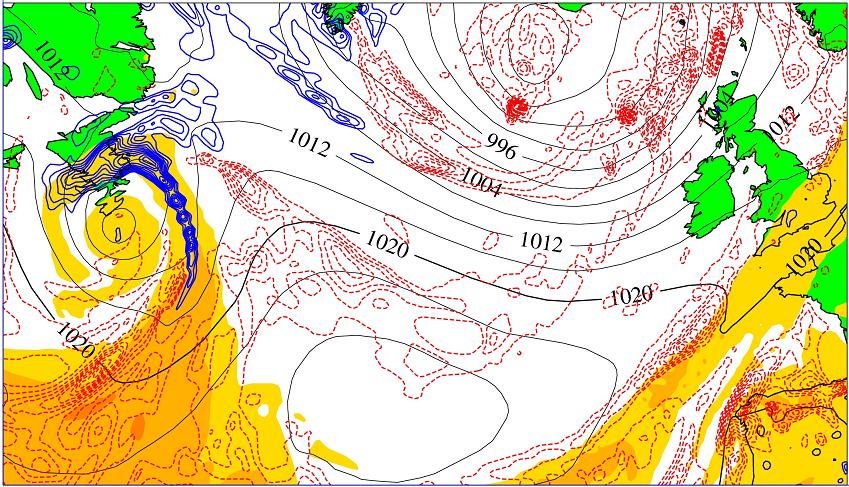}
\includegraphics[width=0.48\linewidth,angle=0,clip=true]{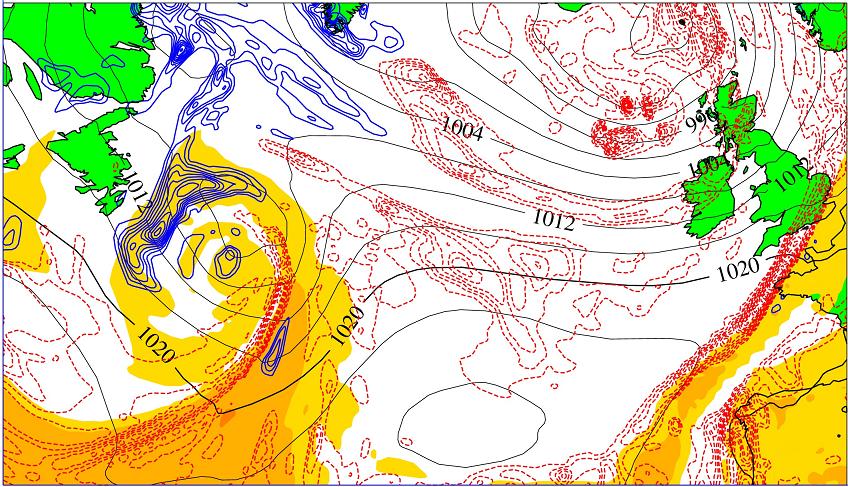}
\\ (e) $PV_e$ / $0$~UTC \hspace{6cm} (f) $PV_e$ / $12$~UTC
\caption{\it \small
Operational output from ARPEGE-IFS model showing analyses and forecasts from $0000$~UTC on 21 September 2011.
The horizontal resolution of the post-processed grid is $0.5^{\circ}$ everywhere, although the resolution of the variable mesh ARPEGE model is about $10$~km over France, Spain and England and about $20$~km over the eastern part of the domain.
MSLP is plotted every $4$~hPa, with a bolder contour for $1020$~hPa.
The potential vorticities are average for the levels $900$, $925$ and $950$~hPa, for
a) $PV_v$ at $0000$~UTC, b) $PV_v$ at $1200$~UTC,
c) $PV_s$ at $0000$~UTC, d) $PV_s$ at $1200$~UTC,
e) $PV_e$ at $0000$~UTC, and f) $PV_e$ at $1200$~UTC. 
Values of $|PV|$ are plotted above the threshold $0.5$~PVUS and then every $0.25$~PVUS intervals.
Positive solid contours are in blue, and negative dashed contours are in red.
Values of $\theta'_w$ at $850$~hPa are represented by (yellow and orange) shaded regions above the threshold $12$, $14$ and $16^{\circ}$C.
\label{Fig_cartes_MSLP_PV_00h_12h}}
\end{figure*}

The moist-air potential vorticity components $PV_v$, $PV_s$ and $PV_e$ are compared in Figures~\ref{Fig_cartes_MSLP_PV_00h_12h}(a)-(f) for a large Atlantic domain, extending from the west of Europe to the Canada.
Values of potential vorticities are averaged for three low levels below $900$~hPa, in order to be less sensitive to numerical problems and to get more significant results.
The shaded regions represent the warm sectors and occluded fronts.
They are delimited by the threshold  $\theta'_w>12^{\circ}$C and with darker shaded areas for higher values of $\theta'_w$.

Solid or dashed lines of equal values of $PV_v$, $PV_s$ and $PV_e$ form coherent signals which are associated with most of the midlatitudes fronts and low-latitudes thermal limits.
These signals do not exhibit random or noisy patterns.
They are advected with the eastward general circulation and they remain close to the same parts of fronts and thermal limits.

% ======================================
% ============ Figure 8 ===================
% ======================================

Charts of $PV_v$ are dominated in (a) and (b)  by positive values close to the fronts (solid lines), with a wide region of weak positive values located over the central North Atlantic.
Charts of $PV_e$ are dominated in (e) and (f)  by negative values (dashed lines), for instance in the central Atlantic, except over warm or occluded fronts where elongated positive structures can be observed.

Charts of $PV_s$ are more balanced in (c) and (d), with the North Atlantic almost free of potential vorticity signal.
Localized and elongated dipole structures are observed close to the cold fronts, with negative values of $PV_s$ on the west side.
The same elongated regions of positive values as the ones observed for $PV_e$ are localized above the warm or occluded fronts.
This suggests an interpretation of these $PV_s$ features in terms of the T-bone structure described in Shapiro and Keyser (1990), with fracturing of the frontal zone near low center.
It would be interesting to analyze further these signals of moderate negative values of $PV_s$, and to determine if they could be associated with intensifying fronts.
For instance, according to forecasters' analyses reported on PRESYG charts  (Santurette and Joly, 2002), the elongated structure located on the central  North Atlantic toward Ireland can be associated with pseudo-warm fronts.

The conclusion of this section is that the descriptions of the meteorological fields in terms of $PV_v$, $PV_s$ or $PV_e$ are not equivalent.
They correspond to different dynamic, thermal and moist aspects of the atmosphere.
It seems that it is easier to analyze the frontal structures of $PV_s$ than those described by $PV_v$ or $PV_e$.
This result may signify that it is indeed important to base the computation of moist-air potential vorticity on the Second Law formulation $c_{pd}\ln(\theta_s)$.

%===============================================
\section{Hints for inversion methods for $PV_s$.} %  (Section 11)
%===============================================
\label{section_inversion_method}

It is well-known that an inversion method requires: 
\begin{itemize}[label=-,leftmargin=3mm,parsep=0cm,itemsep=0cm,topsep=0cm,rightmargin=2mm]
\item  i) an invertible equation, preferably elliptic or no too far hyperbolic; 
\item ii) a relevant set of $N$ equations corresponding to $N$ state variables and; 
\item iii) suitable boundary conditions.
\end{itemize}

The main purpose of the present article is to analyze the diagnostic properties displayed by $\theta_s$ and the associated potential vorticity $PV_s$.
Even though a realistic method may or may not exist to invert $PV_s$, this section and Appendices~C and D analyze the possibility of using $PV_s$ in some way as a starting point for future inversion tools.

A first approach may correspond to the results derived in S01 and S04, where it is demonstrated that $PV_v$ demonstrates a moist-air invertivility principle (although it is not possible to determine all the moist air variables after the inversion process).
It is possible to rely on this property and to control the invertible quantity $PV_v$ expressed as a sum of the specific moist-air entropy version $PV_s$ plus a new component $PV_q$:
\begin{align}
PV_s  & \: = \; PV_v  \: + \: PV_q
  \: , \label{def_PV_s} \\
PV_q  & \: = \; \frac{c_{pd}}{3} \:  PV\!\left[\, \ln(\theta_s/\theta_v) \,\right]
  \: . \label{def_PV_q}
\end{align}
This expression for $PV_q$ results directly from (\ref{def_PV_ln_theta_vl}) and (\ref{def_PV_ln_theta_s}).
It depends on the gradients of $\ln(\theta_{s}/\theta_v)$.
It is explained in Appendix~B that the quotient $\theta_{s}/\theta_v$ mainly depends on the gradients of the water contents $q_t$, $q_l$ and $q_i$, giving the explanation for the notation $PV_{q}$.
The dry-air limit of $PV_q$ is equal to $0$, since  $PV_s = PV_v$ in that case.
The consequence is that $PV_{q}$ observes an  adiabatic (closed) conservative property associated with the join conservation of the dry-air and water contents $q_d$ and $q_t$, as far as the impact of $q_l$ or $q_i$ is small.
A similar component $PV(q_v)$ has been defined and studied in Gao and Zhou (2008).

The reason why it may be easier to manage the sum $PV_s+PV_q$ than $PV_v$ alone is that: 
\begin{itemize}[label=-,leftmargin=3mm,parsep=0cm,itemsep=0cm,topsep=0cm,rightmargin=2mm]
\item i) upper-air patterns of $PV_s$ correspond to the well-known analysis of anomalies of $PV_\theta$ or $PV_v$ located above $600$~hPa, as shown in Figures~\ref{Fig_coupes_PV}(a)-(b); 
\item ii) lower-tropospheric signals can be easier analyzed with  $PV_q$ below $600$~hPa.
\end{itemize}

\begin{figure}[hbt]
\centering
\includegraphics[width=0.49\linewidth,angle=0,clip=true]{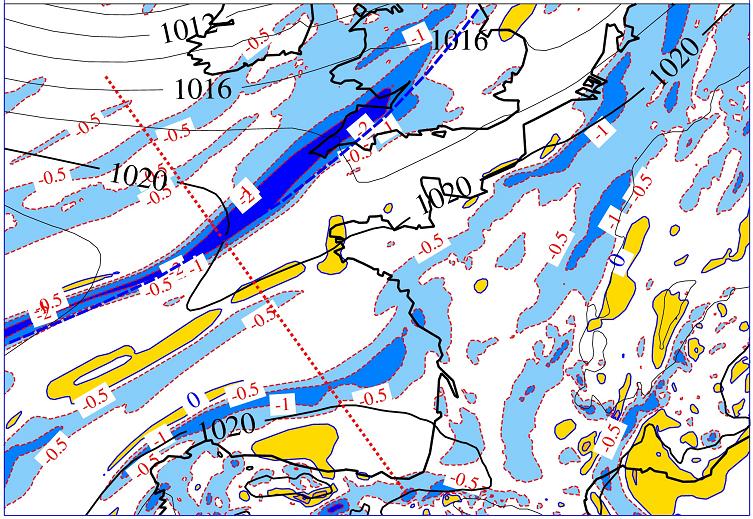}
\includegraphics[width=0.49\linewidth,angle=0,clip=true]{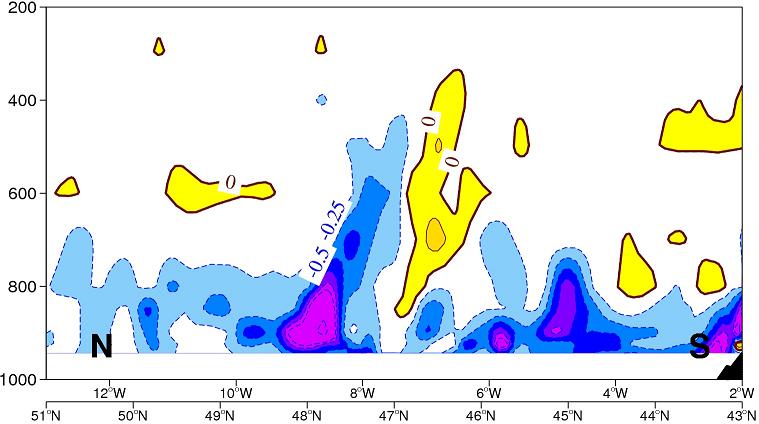}
\\ (a) MSLP and $PV_q$ at $850$~hPa  \hspace{3cm}  (b) Cross-section of $PV_q$
\caption{\it \small
Chart and cross-section as in Figures~\ref{Fig_cartes_PV_850} and \ref{Fig_coupes_PV}, but for the potential vorticity $PV_q$ given by (\ref{def_PV_q}).
Values of $PV_q$ are shaded for positive values (with solid contours) and for negative values (dashed contours)  $\leq-0.5$~PVUS in (a), and $\leq -0.25$~PVUS in (b).
\label{Fig_PVq}}
\end{figure}

The last property can be verified by analyzing the chart of $PV_{q}$ at $850$~hPa shown in Figures~\ref{Fig_PVq}(a) and the cross-section of $PV_{q}$ depicted in Figures~\ref{Fig_PVq}(b).
% ======================================
% ============ Figure 9 ====================
% ======================================
Values of $PV_q$ are negative almost everywhere in (a) and (b).
The limit $-0.25$~PVU extends up to $400$~hPa within the northern limit of the tilted frontal region only, with other significant values of $PV_q$ observed within the lower troposphere only (below $700$~hPa).
A tilted region of positive values are observed close to the southern limit of the front.
It extends up to $300$~hPa close to $7^{\circ}$W and between $46$ and $47^{\circ}$N.

A second approach may folow from inversion processes directly applied to the specific moist-air entropy potential vorticity $PV_s$.
The steps of analysis and  modifications of the $PV_s$ field could rely on the separation observed in Figure~\ref{Fig_cartes_PV_850}(b)  and \ref{Fig_coupes_PV}(b) between upper- troposphere positive values versus lower troposphere negative values, mostly associated with the frontal regions.

The advantage of $PV_s$ with respect of $PV_e$ is that $\theta_s$ observes the Second Law conservative property applied to the specific moist-air entropy, whereas  $\theta_e$ does not, since it is not based on a specific quantity.
The drawback associated with the use of $PV_s$ is that the solenoidal term defined by the term in the right-hand side of the first line of (\ref{def_dPVdt_psi}) does not cancels out with $\psi=\theta_s$, whereas it does with $\psi=\theta_v$.

% ======================================
% ============ Figure 10 ===============
% ======================================
\begin{figure}[hbt]
\centering
\includegraphics[width=0.49\linewidth,angle=0,clip=true]{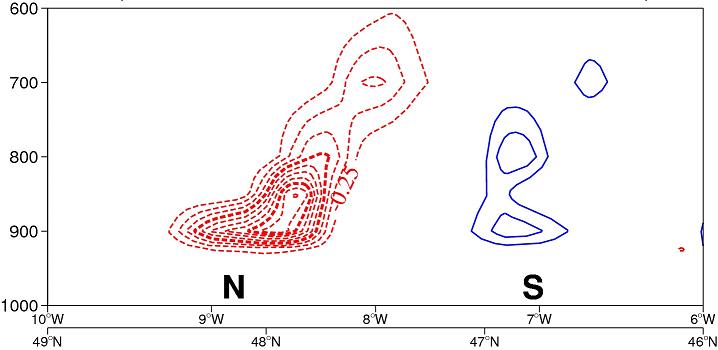}
\includegraphics[width=0.49\linewidth,angle=0,clip=true]{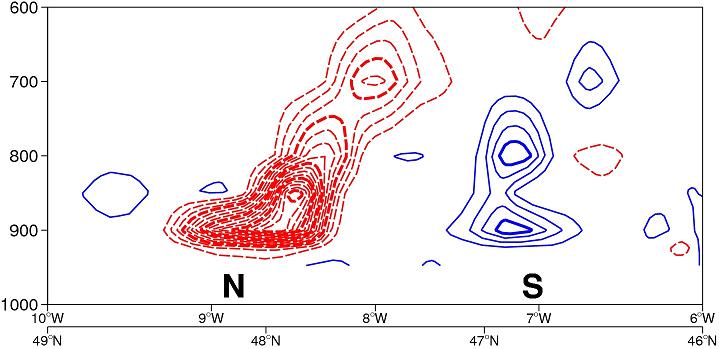}
\\ (a) The solenoidal terms for $\theta_s$ \hspace{3cm} (b) The solenoidal terms for $\theta_e$
\caption{\it \small
Cross-sections as in Figures~\ref{Fig_coupe_PV_dsdp}, but for the solenoidal terms (units  $10^{-10}$~m~s${}^{-4}$~K${}^{-1}$) for 
(a) $\theta_s$, and
(b) $\theta_e$.
Contours are shown at $0.25$~unit intervals, with solid (blue) contours $\geq 0.25$ and dashed (red) contours $\leq -0.25$.
\label{Fig_Solenoidal}}
\end{figure}
However, the same problem exist in inversion methods applied to $PV_e$, with $\psi=\theta_e$ generating a non-cancelling solenoidal term.
Numerical values of the two solenoidal terms generated by $\psi=\theta_s$ or $\theta_e$ are compared in Figures~\ref{Fig_Solenoidal}(a,b).

The promising result is that the solenoidal term obtained with $\theta_s$ is about half that obtained with $\theta_e$.
It has been confirmed (not shown) that the solenoidal terms obtained with $\theta$ and $\theta_v$ are, as expected, very small in comparison with that obtained with $\theta_s$ ($10$ times lower with $\theta$ and $1000$ times lower with $\theta_v$).

The problem generated by the non-vanishing solenoidal term is thus less important with $PV_s$ than with $PV_e$.
The problem of the non-elliptic character for the inversion operator may also be less important with the observed moderate negative values of $PV_s$, in comparison with the larger negative values for $PV_e$.

%========================
\section{Conclusions.} %  (Section 12)
%========================
\label{section_conclude}

The purpose of this article was to define a  moist potential vorticity $PV_s$ written in terms of the specific moist entropy and the potential temperature $\theta_s$ described in M11.
An associated new $PV$ unit has been derived, in order to recover the same numerical values of about $1.5$~PVUS at the tropopause, where dynamical anomalies are classically analyzed with Ertel's component $PV_{\theta}$.

Outputs from a NWP model are used to illustrate and analyze the spatial and temporal patterns of the new moist-air potential vorticity $PV_s$.
The other formulations $PV_v$ and $PV_e$ obtained with the well-known virtual and equivalent potential temperatures $\theta_v$ and $\theta_e$ are compared with $PV_s$.

Analyses of isobaric charts and vertical cross-sections show that: 
\begin{itemize}[label=-,leftmargin=3mm,parsep=0cm,itemsep=0cm,topsep=0cm,rightmargin=2mm]
\item i) the upper level  positive anomalies of $PV_s$ are similar to the well-known Ertel's $PV_{\theta}$ structures; 
\item ii) the low-level features are associated with frontal regions and with moderate negative values (about half the values of $PV_e$);
\item iii) the low level properties displayed by $PV_s$ close to the fronts might be associated with the criteria which control conditional symmetric instability and slantwise convection;
\item iv) the low level solenoidal term generated by $PV_s$ is half the one generated by  $PV_e$.
\end{itemize}

The main justification for the use of $PV_s$ in place of $PV_e$ is that the conservation principle associated with the Second Law must be expressed with material derivatives of a specific moist-air quantity.
It is explained that neither $\theta'_w$ nor $\theta_e$ fulfils this requirement, because they are associated with pseudo-adiabatic processes expressed by unit mass of dry air, not with specific values.
One could say that ``$\theta'_w$ and $\theta_e$ are pseudo-adiabatically conserved quantities which are not conservative, whereas $\theta_s$ is the conservative variable associated with the specific moist air entropy which is conserved for closed, adiabatic moist-air processes''.
 
It is possible to define a water potential vorticity $PV_q$ computed as the difference between $PV_s$ and the virtual value $PV_v$.
It is shown that spatial structures of low-level negative values of $PV_s$ are highlighted by $PV_q$, with larger negative values which may be used to analyze, modify and control those low-level signals which are more difficult to see with $PV_s$, and are even absent in $PV_v$.

The main interest for introducing the water component $PV_q$ would be that is may allow a partition of the invertible component $PV_v$ into a weighting sum of the Ertel's $PV$ and $PV_q$, each of them controlling different, separated parts of the atmosphere.
This partition may be a relevant one, since the virtual potential vorticity $PV_v$ seems to be the only moist air $PV$ which demonstrates an invertibility principle.

It could be interesting to test alternative inversion methods based on $PV_s$ alone, with the positive upper-level and negative lower-level structures which could be analyzed, modified and controlled with the same horizontal charts and vertical cross-sections of $PV_s$ as shown in the paper.

More sophisticated numerical modelling and observational studies of $PV_s$ are required to assess the possibility of a realistic inversion tool based on $PV_s$ alone, or on the pair of potential vorticities $PV_{\theta}$ and $PV_q$.

As a final perspective, it may be possible to improve the variational moist inversion method described by Arbogast {\it et al.\/} (2008) and based on the ideas of Vallis (1996), by testing for instance a moist total available enthalpy norm to solve the corresponding variational problem, with the moist entropy used in the present paper entering the moist norm.
It will be necessary to derive an associated moist quasi-geostrophic set of equations, with prognostic equations for water species which might correspond to the moist entropy and water components $PV_s$ and $PV_q$ defined here.

%-------------------------------------------------------------------------
%    Acknowledgement(s) : \acknowledgement or \acknowledgements
%-------------------------------------------------------------------------

%%%%%%%%%%%%%%%%%%%%
% ACKNOWLEDGEMENTS %
%%%%%%%%%%%%%%%%%%%%%%%%%%%%%%%%%%%%%%%%%%%%%%%%%%%%%%%%
% The title "Acknowledgement"  is provided by : \ack   %
% The title "Acknowledgements" is provided by : \acks  %
%%%%%%%%%%%%%%%%%%%%%%%%%%%%%%%%%%%%%%%%%%%%%%%%%%%%%%%%

\vspace{5mm}
\noindent{\large\bf Acknowledgements}
\vspace{2mm}

The author is most grateful to Patrick Santurette, Jean-Fran\c{c}ois Geleyn, Etienne Blot and Philippe Arbogast for stimulating discussions.
The author would like to thank the anonymous referees for their constructive comments, which helped to improve the manuscript.

%-----------------
%   APPENDIXES
%-----------------

%----------------- -------------------------------------
%    APPENDIX - A    
%----------------- -------------------------------------
%----------------------------------------------------------------------
\vspace{4mm}
\noindent
{\bf Appendix A. List of symbols and acronyms.}
%----------------------------------------------------------------------
             \label{appendixSymbol}
\renewcommand{\theequation}{A.\arabic{equation}}
  \renewcommand{\thefigure}{A.\arabic{figure}}
   \renewcommand{\thetable}{A.\arabic{table}}
      \setcounter{equation}{0}
        \setcounter{figure}{0}
         \setcounter{table}{0}
\vspace{1mm}
\hrule

\begin{tabbing}
 -------------\=  --------------------------------------- --\= \kill
 $\alpha$ \> $=1/\rho$ the specific volume \\
 $c_{pd}$ \> specific heat for dry air   \>($1004.7$~J~K${}^{-1}$~kg${}^{-1}$) \\
 $c_{pv}$ \> spec. heat for water vapour \>($1846.1$~J~K${}^{-1}$~kg${}^{-1}$) \\
 $c_{l}$  \> spec. heat for liquid water \>($4218$~J~K${}^{-1}$~kg${}^{-1}$) \\
 $c_{i}$  \> spec. heat for ice water    \>($2106$~J~K${}^{-1}$~kg${}^{-1}$) \\
 $c_p$ \> specific heat at constant pressure for moist air, \\
       \> $ = \: q_d \: c_{pd} + q_v \: c_{pv} + q_l \: c_l  + q_i \: c_i $ \:
          $ = \: q_d \: ( \: c_{pd} + r_v \: c_{pv} + r_l \: c_ l + r_i \: c_ i)$ \\
 $C_1$, $C_2$ \> two terms appearing in a  moist entropy equation \\
 $c_p^{\ast}$ \> a specific heat depending on $r_t$\\
 $\delta$ \> $=R_v/R_d-1 \approx 0.608$ \\
 $\eta$   \> $=1+\delta =R_v/R_d \approx 1.608$ \\
 $\kappa$ \> $=R_d/c_{pd}\approx 0.2857$ \\
 $\gamma$ \> $= \eta \: \kappa \ = R_v/c_{pd} \approx 0.46$ \\
 $\lambda$ \> $= c_{pv}/c_{pd}-1 \approx 0.8375$ \\
 $e_r$      \> the water vapour reference partial pressure,
               with $\: e_r = e_{ws}(T_0) \approx 6.11$~hPa \\
 $f$ \> the Coriolis parameter \\
 ${\Lambda}_r$ \> $= [ (s_{v})_r - (s_{d})_r ] / c_{pd} \approx 5.87$ \\
 $L_v (T)$ \> $=h_v-h_l$: Latent heat of vaporisation \\
 $L_s (T)$ \> $=h_v-h_i$: Latent heat of sublimation \\
 $L_v (T_0)$ \> $= 2.501$~$10^{6}$~J~kg${}^{-1}$ \\
 $L_s (T_0)$ \> $= 2.835$~$10^{6}$~J~kg${}^{-1}$ \\
 $N^2_s$ \>  a moist-air squared Brunt-V\"{a}is\"{a}l\"{a} frequency \\
 $p$      \> $=p_d + e$: local value for the pressure \\
 $p_r$  \> $=(p_d)_r + e_r$: reference pressure ($p_r=p_0$)\\
 $p_d$     \> local dry-air partial pressure \\
 $(p_d)_r$ \> reference dry-air partial pressure ($\equiv p_r-e_r$)\\
 $p_0$     \> $=1000$~hPa: conventional pressure \\
 $\psi$    \> dummy scalar value \\
 $PV(\theta)$   \> Ertel's potential vorticity (PVU) \\
 $PV_{\theta}$   \> dry-air entropy potential vorticity (PVUS) \\
 $PV_v$   \> virtual potential vorticity (PVUS) \\
 $PV_e$   \> equivalent potential vorticity (PVUS) \\
 $PV_{es}$   \> saturation equivalent potential vorticity  (PVUS) \\
 $PV_s$   \> moist-air entropy potential vorticity (PVUS) \\
 $PV_q$   \> water potential vorticity (PVUS) \\
 $PV_{s1}$   \> approximate version for $PV_{s}$  (PVUS) \\
 $PV_{q1}$   \> approximate version for $PV_{q}$ (PVUS) \\
 $PV_{l1}$   \> liquid-water potential vorticity (PVUS) \\
 $q_{sw}$   \> saturated vapour content over liquid water \\
 $q_{si}$   \> saturated vapour content over ice water \\
 $q_{w}$   \> wet bulb potential vorticity \\
 $q_{g}$   \> entropic potential vorticity \\
 $q_{ge}$   \> equivalent potential vorticity \\
 $q_{d}$   \> $={\rho}_d / {\rho}$: specific content for dry air \\
 $q_{v}$   \> $={\rho}_v / {\rho}$: specific content for water vapour \\
 $q_{l}$   \> $={\rho}_l / {\rho}$: specific content for liquid water \\
 $q_{i}$   \> $={\rho}_i / {\rho}$: specific content for ice water\\
 $q_t  $   \> $= q_v+q_l+q_i$: total specific content of water \\
 $r_{v}$   \> $=q_{v}/q_{d}$: mixing ratio for water vapour \\
 $r_{sw}$   \> saturated vapour mixing ratio over liquid water\\
 $r_{r}$   \> reference mixing ratio for water species 
            ($\eta\:r_{r} \equiv e_r / (p_d)_r$ and 
             $r_{r} \approx 3.82$~g~kg${}^{-1}$) \\
 ${\rho}_d$   \> specific mass for  dry air  \\
 ${\rho}_v$   \> specific mass for  water vapour \\
 ${\rho}_l$   \> specific mass for  liquid water \\
 ${\rho}_i$   \> specific mass for  ice water \\
 ${\rho}$     \> specific mass for  moist air  $={\rho}_d+{\rho}_v+{\rho}_l+{\rho}_i$  \\
 $R_v$   \> water vapour gas constant ($461.52$~J~K${}^{-1}$~kg${}^{-1}$) \\
 $R_d$   \> dry-air gas constant      ($287.06$~J~K${}^{-1}$~kg${}^{-1}$) \\
 $R$     \> $ = q_d \: R_d + q_v \: R_v$: gas constant for moist air $ R = q_d \:(\: R_d + r_v \: R_v)$ \\
 $s$       \> specific moist entropy \\
 ${s}_{ref}$   \> a reference specific entropy  \\
 $(s_{d})_r$  \>  reference values for the entropy of dry air 
                  at $T_r$ and $(p_d)_r$: $6777$~J~K${}^{-1}$~kg${}^{-1}$\\
 $(s_{v})_r$  \> reference values for the entropy of water vapour
                  at $T_r$ and $e_r$: $12673$~J~K${}^{-1}$~kg${}^{-1}$\\
 $s_e$       \> a specific moist equivalent entropies \\
 $s^{\ast}_r$  \> a reference moist entropy depending on $r_t$ \\
 $S$       \> a specific moist entropy (Hauf and H\"{o}ller, 1987) \\
 $T$       \> local temperature \\
 $T_{r}$   \> the reference temperature ($T_r\equiv T_0$) \\
 $T_{0}$   \> zero Celsius temperature ($=273.15$~K) \\
 $\theta$         \> $ = T\:(p_0/p)^{\kappa}$: the (dry-air) potential temperature\\
 ${\theta}_{v}$   \> the virtual potential temperature \\
 ${\theta}_{l}$   \> the liquid-water potential temperature \\
 ${\theta}_{e}$   \> the equivalent potential temperature\\
 ${\theta}_{es}$   \> the saturation equivalent potential temperature\\
 ${\theta}^{\star}_{e}$   \> a saturation equivalent potential temperature  (Schubert 2004)\\
 ${\theta}_{S}$   \> a moist entropy potential temperature  (Hauf and H\"{o}ller, 1987) \\
 ${\theta}_{s}$   \> the moist entropy potential temperature (M11) \\
 $({\theta}_{s})_1$   \> approximate version of ${\theta}_{s}$ \\
 $\boldsymbol{\zeta}_a$ \> the absolute vorticity 3D vector \\
 $\boldsymbol{\nabla}$ \> gradient 3D operator \\
 $\boldsymbol{\Omega}$ \> angular velocity of the Earth (3D vector) \\
 $\boldsymbol{u}$ \> the velocity (3D vector) \\
 $\boldsymbol{F}$ \> frictional force 3D vector
\end{tabbing}

\vspace{1mm}
\hrule

\vspace{3mm}
%----------------------------------------------------------------------
\noindent
{\bf Appendix~B. The approximations $PV_{s1}$ and $ PV_{q1}$.}
%----------------------------------------------------------------------
             \label{AppB_PVs1_PVq1}
\renewcommand{\theequation}{B.\arabic{equation}}
  \renewcommand{\thefigure}{B.\arabic{figure}}
   \renewcommand{\thetable}{B.\arabic{table}}
      \setcounter{equation}{0}
        \setcounter{figure}{0}
         \setcounter{table}{0}
\vspace{1mm}

Since it is possible to approximate ${\theta}_{s}$ by $({\theta}_{s})_1$, it is worthwhile to determine to what extent $PV_s$ and $PV_q$ may be modified by this approximation.
The hope is to obtain simple analytic expressions for $PV_s$ and $PV_q$ which could be easier to manage and possibly to derive easier physical interpretations.

Let us approximate the moist entropy potential temperature ${\theta}_{s}$ by $({\theta}_{s})_1$ given by (\ref{def_theta_s1}).
The  analog of  (\ref{def_PV_ln_theta_s}) and (\ref{def_PV_q}) is the exact separation of $PV_{s1}$ into the sum of
\begin{align}
PV_{s1}  & \: = \; \frac{c_{pd}}{3} \:  PV\!\left[\, \ln\{(\theta_s)_1\} \,\right]
  \: , \label{def_PV_s1a} \\
PV_{q1}  & \: = \; \frac{c_{pd}}{3} \:  PV\!\left[\, \ln\{(\theta_s)_1/\theta_{v}\} \,\right]
  \: . \label{def_PV_q1}
\end{align}
The component $PV_{q1}$ exactly represent the difference between the approximate entropy component $PV_{s1}$ and the invertible one $PV_{v}$.

From the definitions (\ref{def_theta_s1})  for $(\theta_s)_1$ and (\ref{def_PV_ln_theta}) for $PV_{\theta}$, the moist entropy component $PV_{s1}$ can be written as
\begin{align}
\! \! PV_{s1}  & =
    PV_{\theta}
     +  \frac{c_{pd}}{3} \;
    PV\!\left( {\Lambda}_r\:q_t  - \frac{L_v\:q_l + L_s\:q_i}{{c}_{pd}\:T} \right)
  . \label{def_PV_s1_ter}
\end{align}

The water content component $PV_{q1} = PV_{s1} - PV_{v}$ can be computed as the difference (\ref{def_PV_s1_ter}) minus (\ref{def_PV_vl_bis}), or directly with the quantity $\ln\{(\theta_s)_1 / \theta_{v}\}$ evaluated from (\ref{def_Theta_v1}) and (\ref{def_theta_s1}) and with the Ertel's potential vorticity $PV_{\theta}$ which cancels out between (\ref{def_PV_vl_bis}) and (\ref{def_PV_s1_ter}), leading to the result
\begin{align}
PV_{q1}  & \: = \; 
 - \: \frac{c_{pd}}{3} \; 
 \frac{PV\!\left( \delta\: q_v  -  q_l  - q_i \right)}
      {\left( 1  + \delta\: q_v  - q_l  - q_i \right)}
 \; + \: \frac{c_{pd}}{3} \;\: 
PV\!\left(\: {\Lambda}_r\:q_t  \: - \: \frac{L_v\:q_l + L_s\:q_i}{{c}_{pd}\:T} \:\right)
  \: . \label{def_PV_q1_bis}
\end{align}
This expression mainly depends on the gradients of water contents  $q_v$, $q_l$ and $q_i$, with small impact due to variations of $L_v(T)/T$ and $L_s(T)/T$ with $T$ in the last $PV$ term of (\ref{def_PV_q1_bis}).

It is possible to express $PV_v$,  $PV_{s1}$  and $PV_{q1}$, given by (\ref{def_PV_vl_bis}),  (\ref{def_PV_s1_ter}) and (\ref{def_PV_q1_bis}),  in terms of the variables $\theta$,  $q_t$, $q_l$ and $q_i$, with $q_v$ expressed as $q_t-q_l-q_i$.
The resulting moist $PV$  components are written as 
\begin{align}
PV_v  
   & \: = \; 
 PV_{\theta}
 \; + \: 
 \frac{c_{pd}}{3} \; 
  \delta\: 
   \frac{\theta}{\theta_v} 
PV\!\left( q_t \right)
 \mbox{\hspace{5mm}}
 \; -  \:  
   \frac{c_{pd}}{3} \;
   \left(1+\delta\right) \:
    \frac{\theta}{\theta_v} \:
  \left[ \:
    PV\!\left( q_l \right)
   +
    PV\!\left( q_i \right)
  \: \right]
\: , \label{def_PV_v_qt} \\
PV_{s1}  
   & \: = \; 
 PV_{\theta}
 \; + \: 
 \frac{c_{pd}}{3} \; {\Lambda}_r \:
PV\!\left( q_t \right)
 \mbox{\hspace{7mm}}
 \; - PV\!\left(  \frac{L_v\: q_l}{3\:T} \right)
 \: - PV\!\left(  \frac{L_s\: q_i}{3\:T} \right)
\: , \label{def_PV_s1_qt} \\
PV_{q1}  
   & \: = \; 
 \frac{c_{pd}}{3} \; 
\left(\:
{\Lambda}_r
-
  \delta\: 
   \frac{\theta}{\theta_v} 
\:\right)
PV\!\left( q_t \right)
 \mbox{\hspace{1mm}}
 \; - PV\!\left(  \frac{L_v\: q_l}{3\:T} \right)
 \: - PV\!\left(  \frac{L_s\: q_i}{3\:T} \right)
\nonumber \\
 & \;
 \mbox{\hspace{52mm}}
\; +  \:  
   \frac{c_{pd}}{3} \;
   \left(1+\delta\right) \:
    \frac{\theta}{\theta_v} \:
  \left[ \:
    PV\!\left( q_l \right)
   +
    PV\!\left( q_i \right)
  \: \right]
\: . \label{def_PV_q1_qt}
\end{align}
The first terms on the right-hand sides of  (\ref{def_PV_v_qt})-(\ref{def_PV_q1_qt}) depend on $PV(\theta)$ and $PV(q_t)$, whereas the other terms depends on the potential vorticities of condensed water $PV(q_l)$ or $PV(q_i)$.

\vspace{3mm}
%----------------------------------------------------------------------
\noindent
{\bf Appendix~C. An inversion method for $PV_{s1}$ and $PV_{q1}$: the non-saturated moist air.}
%----------------------------------------------------------------------
             \label{AppC_Inversion_nonsat}
\renewcommand{\theequation}{C.\arabic{equation}}
  \renewcommand{\thefigure}{C.\arabic{figure}}
   \renewcommand{\thetable}{C.\arabic{table}}
      \setcounter{equation}{0}
        \setcounter{figure}{0}
         \setcounter{table}{0}
\vspace{1mm}

The aim of this Appendix is to make more explicit the second approach suggested at the end of section~\ref{section_inversion_method}, for the special case of non-saturated moist air.

The interest of an inversion process to be applied to $PV_v$ relies on the capacity to modify this field $PV_v$ in a relevant way before the inversion step.
The method is currently used to manage the upper-troposphere and  lower-stratosphere anomalies of $PV_{\theta}$, which are almost the same as those of $PV_v$ in these dry regions.

The problem is different for the low-level $PV_v$ informations, which is clearly not associated with frontal limits, as shown in Sections~\ref{section_PV_charts}, \ref{section_PV_cross} and \ref{section_PV_series}.
The alternative method is to diagnose more clearly the low-level information with the help of the water component $PV_{q1}$.
Comparisons of Figures~\ref{Fig_coupes_PV}(b) and \ref{Fig_PVq}(b) show that almost the same regions of negative values can be diagnosed with $PV_{s}$  or $PV_{q}$, except that the threshold $0$~PVUS for $PV_{s}$ must be replaced by $-0.5$~PVUS for $PV_{q}$.

Accordingly, the aim is to express $PV_v$ as a weighting sum of $PV_{\theta}$ and $PV_{q1}$, with $PV_{\theta}$ suitable for upper-level analysis and $PV_{q1}$ for lower-level analysis.
The non-saturated moist-air version is considered in this section and the computations valid for the saturated moist air will be conducted in the next section.
The non-saturated and dry-air cases are  different, with {\it a priori\/} a large impact of $0 < q_v < q_{sw}$ on the computations of  $(\theta_s)_1$, $PV_{s1}$ and $PV_{q1}$.

The non-saturated version of $PV_v$ and $PV_{q1}$ are obtained from  (\ref{def_PV_v_qt}) and (\ref{def_PV_q1_qt})  for $q_l = q_i = 0$ and for $q_t=q_v$.
The result is
\begin{align}
PV_{v}  & = \: 
   PV_{\theta}
     \; + \: 
   \frac{c_{pd}}{3}
         \; {\delta}_v
   \;  PV ( q_v  )
  \: , \label{def_PV_v_app2} \\
PV_{q1}  & = \:  
      \frac{c_{pd}}{3} \; 
      \left(\:
         {\Lambda}_r
              -
         {\delta}_v 
     \:\right)
      PV ( q_v  )
  \: . \label{def_PV_q1_app2}
\end{align}
The term ${\delta}_v$ is equal to $\delta\:(\theta/\theta_v) = \delta\:/(1+\delta\:q_v)$ in non-saturated conditions.
The next step is to compute $PV(q_v)$ from (\ref{def_PV_q1_app2}) and to put the result into (\ref{def_PV_v_app2}).
The result is $PV_v$ expressed as the weighting sum 
\begin{align}
PV_{v} 
    & = \;
     PV_{\theta}
   \; + \: 
     \left( 
      \frac{{\delta}_v}{ {\Lambda}_r \: - {\delta}_v}
     \right) \;
     PV_{q1}
  \: . \label{def_PV_v_app3}
\end{align}
The property (\ref{def_PV_v_app3}) explains how it is possible to take into account any modified version of the two components $PV_{\theta}$ and $PV_{q1}$ and how they can be recombined, with the coefficient ${\delta}_v  /({\Lambda}_r \: - {\delta}_v)$ acting on the water component, to give an updated value for $PV_{v}$ which may enter an inversion process.

Since ${\delta} = 0.608$ and because $q_v$ is a small term in the atmosphere, values of ${\delta}_v$ remain close to ${\delta}$.
The impact of $q_v$ on $\delta_v$ is of the order of $100 \times q_v$ (in per cent), with for instance ${\delta}_v = 0.6$ for $q_v=20$~g~kg${}^{-1}$.
The coefficients acting to mix the components $PV_{\theta}$ and $PV_{q1}$ in (\ref{def_PV_v_app3}) can thus be evaluated with ${\delta}_v \approx  0.608$ and ${\Lambda}_r = 5.87$ to give ${\delta}_v  / ({\Lambda}_r \: - {\delta}_v) \approx 0.116$.
Even for a large value of $q_v=20$~g~kg${}^{-1}$ the coefficients is almost the same: $0.114$.
A mean value of  $0.115$ for ${\delta}_v  / ({\Lambda}_r \: - {\delta}_v)$  leads to the result
\begin{align}
PV_{v} 
    &  \;  \approx \;
     PV_{\theta}
   \; + \: 
     0.115 \;
     PV_{q1}
  \: . \label{def_PV_v_app4}
\end{align}

The consequence of (\ref{def_PV_v_app4}) is that a change of $1$~PVUS in $PV_{q1}$ implies a small change of $0.1$~PVUS in the virtual component  $PV_v$.
This means that the modified values of $PV_v$ would always remain relevant ones, even if $PV_q$ is largely modified.

The term ${\Lambda}_r$  is a consequence of the Third Law.
It  has a  large impact on the definition of $ PV_{q1}$ given by (\ref{def_PV_q1_app2}) and on the associated coefficient ${\delta}_v  / ({\Lambda}_r \: - {\delta}_v)$ in (\ref{def_PV_v_app3}).

The value ${\delta}_v \approx  0.608$ is not allowed for ${\Lambda}_r$, because this value makes the coefficient infinite.
According to M11, this value cannot be attained by the range of $4.67$ to $6.47$ for ${\Lambda}_r$, depending on the values of $T_r$ and $p_r$.
The usual choice of a null standard entropy for dry air and liquid water corresponds to ${\Lambda}_r \approx L_v\:/(c_{pd}\:T) \approx 10$ and to $PV_{s1} \approx PV_e$.
The other choice of null standard  entropies for dry air and water vapour corresponds to ${\Lambda}_r = 0$  and to the trivial result $PV_{s1}=PV_{\theta}$.

\vspace{3mm}
%----------------------------------------------------------------------
\noindent
{\bf Appendix~D. An inversion method for $PV_{s1}$ and $PV_{q1}$: the saturated moist air..}
%----------------------------------------------------------------------
             \label{AppD_Inversion_sat}
\renewcommand{\theequation}{D.\arabic{equation}}
  \renewcommand{\thefigure}{D.\arabic{figure}}
   \renewcommand{\thetable}{D.\arabic{table}}
      \setcounter{equation}{0}
        \setcounter{figure}{0}
         \setcounter{table}{0}
\vspace{1mm}

The results obtained in Appendix~C for the non-saturated moist air is generalized in this section for the case of a saturated moist air.
For the sake of simplicity, the computations will be limited to the description of liquid water content, with the properties $q_v=q_{sw}$ and $q_l=q_t-q_{sw}$.
The formulae verified when liquid water is replaced by ice water will be obtained by replacing $q_{sw}$ by $q_{si}$ and $L_v$ by $L_s$ in the next saturated liquid-water $PV$ formulations.

If liquid water content exists, the virtual and moist entropy saturated components (\ref{def_PV_v_qt}) and  (\ref{def_PV_s1_qt}) write
\begin{align}
PV_{v}  & = \: 
   PV_{\theta}
     \; + \: 
   \frac{c_{pd}}{3}
         \; {\delta}_v
   \;  PV\!\left( q_t  \right)
   \; - \: 
   \frac{c_{pd}}{3} \;
     \left[ 
   \frac{{\delta}_v \:(1+\delta)}{\delta}
     \right]
    PV\!\left( q_l  \right)
  \: , \label{def_PV_v_sat_app2} \\
PV_{s1}  & = \:
   PV_{\theta} 
     \; + \: 
   \frac{c_{pd}}{3}
   \; {\Lambda}_r \;
       PV\!\left( q_t \right)
   \; - \: 
   \frac{c_{pd}}{3} \;
       PV\!\left( 
          \frac{L_v \: q_l}{{c}_{pd}\:T}
          \right)
  \: , \label{def_PV_s1_sat_app2} 
\end{align}
where ${\delta}_v = {\delta} \: (\theta/\theta_v)$.

The differences between the non-saturated and the saturated versions  concern only the last $PV$ terms of  (\ref{def_PV_v_sat_app2}) and (\ref{def_PV_s1_sat_app2}), which mainly depend on the gradients of $q_l$, because the term $\:q_l\:PV[\:L_v(T)/T\:]$ is {\it a priori\/}  much smaller than the term $[\:L_v(T)/T\:]\:PV(q_l)$.

The difference between (\ref{def_PV_s1_sat_app2}) and (\ref{def_PV_v_sat_app2}) generates the water component
\begin{align}
PV_{q1}  & = \:
   \frac{c_{pd}}{3}
      \left(\:
         {\Lambda}_r
              -
         {\delta}_v 
     \:\right) \:
   PV\!\left( q_t  \right)
\; + \: 
   \frac{c_{pd}}{3} \;
     \left[ 
   \frac{{\delta}_v \:(1+\delta)}{\delta}
     \right]
   PV\!\left( q_l  \right)
 \; - \:
   \frac{c_{pd}}{3} \;
   \;  PV\!\left( 
          \frac{L_v \: q_l}{{c}_{pd}\:T}
          \right)
  \: . \label{def_PV_q1_sat_app2} 
\end{align}

If ${\Lambda}_r \neq {\delta}_v$, it is then possible to express $PV(q_t)$ from (\ref{def_PV_q1_sat_app2}) in terms of $PV_{q1}$ plus additional terms mainly depending on the gradients of $q_l$ (if the term depending on $PV[\:L_v(T)/T\:]$ is neglected).
This expression for $PV(q_t)$ can be replaced in (\ref{def_PV_v_sat_app2}) and (\ref{def_PV_s1_sat_app2}), leading to the result
\begin{align}
PV_{v} 
    & \; = \;
     PV_{\theta} \;
   \; + \: 
     \left( 
      \frac{{\delta}_v}{ {\Lambda}_r \: - {\delta}_v}
     \right)
     PV_{q1} \;
    \; + \:
     PV_{l1}
  \: , \label{def_PV_v_sat_app3}
\end{align}
where the liquid-water component is equal to
\begin{align}
PV_{l1}  
    & \; = 
   \;
   \frac{c_{pd}}{3} \:
     \left( 
      \frac{{\delta}_v}{ {\Lambda}_r \: - {\delta}_v}
     \right)
     PV\!\left( 
          \frac{L_v \: q_l}{{c}_{pd}\:T}
          \right)  
  \; - \; 
   \frac{c_{pd}}{3} \:
     \left( 
      \frac{{\Lambda}_r}{ {\Lambda}_r \: - {\delta}_v}
     \right)
     \left[ 
   \frac{{\delta}_v \:(1+\delta)}{\delta}
     \right]
     PV\!\left( q_l  \right)
 . \label{def_PV_l1_sat}
\end{align}
Comparison of (\ref{def_PV_v_sat_app3}) with the non-saturated result (\ref{def_PV_v_app3})  shows that the terms depending on $PV_{\theta}$ and $PV_{q1}$ are the same, except that ${\delta}_v = \delta\:/(1+\delta\:q_v+q_l)$ depends on $q_l$.
The new liquid-water component $PV_{l1}$ must be added to $PV_{v}$ which mainly depends on $PV(q_l)$.
The non-saturated result is easily obtained if $q_l=0$, with ${\delta}_v = \delta\:/(1+\delta\:q_v)$ and with $PV_{l1}$ which exactly cancels out.

An approximate method can be imagined to manage the water component $PV_{q1}$ and the liquid-water component $PV_{l1}$ in  (\ref{def_PV_v_sat_app3}).

The terms depending on $q_l$ might be discarded in the definition (\ref{def_PV_q1_sat_app2}) for $PV_{q1}$, with accordingly $PV_{l1}$ removed from the saturated result (\ref{def_PV_v_sat_app3}).
The simplified results  corresponds to a ``just-saturated'' moist air version for  (\ref{def_PV_v_sat_app3}).
Interestingly, it is the same as the non-saturated system (\ref{def_PV_v_app3}),  with $q_v$ replaced by $q_t$.

The main advantages of this ``just-saturated'' assumption is that the same formulation could be used for non-saturated and saturated atmospheric conditions, and that the two components $PV_{\theta}$ and $PV_{q1}$ are probably sufficient for analyzing and modifying $PV_v$.
Additional validations are required to determine whether or not the large scale structures of the liquid-water component $PV_{l1}$ are indeed small correction terms.

%\end{document}

%-------------------------------------------------------------------------
%    REFERENCES
%-------------------------------------------------------------------------

%\begin{thebibliography}

%\newpage

\vspace{5mm}
\noindent{\large\bf References}
\vspace{2mm}

\noindent{$\bullet$ Arbogast P, Maynard K, Crepin F.} {2008}.
{Ertel potential vorticity inversion using a digital filter initialization method.
{\it Q. J. R. Meteorol. Soc.}
{\bf 134} (634):
1287--1296.}

\noindent{$\bullet$ Bennetts DA, Hoskins BJ.} {1979}.
{Conditional symmetric instability. A possible explanation for frontal rainbands.
{\it Q. J. R. Meteorol. Soc.}
{\bf 105} (446):
945--962. (BH79)}

\noindent{$\bullet$ Betts AK.} {1973}.
{Non-precipitating cumulus convection and its parameterization.
{\it Q. J. R. Meteorol. Soc.}
{\bf 99} (419):
178--196.}

\noindent{$\bullet$ Bolton D.} {1980}.
{The computation of Equivalent Potential Temperature.
{\it Mon. Weather Rev.}
{\bf 108,} (7):
1046--1053.}

\noindent{$\bullet$ Davies HC, Wernli H} {1997}.
{On studying the structure of synoptic systems.
{\it Meteorol. Appl.}
{\bf 14,} (4):
365--374.}

\noindent{$\bullet$ Emanuel KA.} {1979}.
{Inertial Instability and Mesoscale Convective Systems. 
Part I: Linear Theory of Inertial Instability in Rotating 
Viscous Fluids.
{\it J. Atmos. Sci.}
{\bf 36} (12):
2425--2449.}

\noindent{$\bullet$ Emanuel KA.} {1983}.
{On Assessing Local Conditional Symmetric Instability 
from Atmospheric Soundings.
{\it Mon. Wea. Rev.}
{\bf 111} (10):
2016--2033.}

\noindent{$\bullet$ Emanuel KA, Fantini M, Thorpe A.} {1987}.
{Baroclinic instability in an environment of small stability 
to slantwise moist convection. part I: two-dimensional 
models.
{\it J. Atmos. Sci.}
{\bf 44} (12):
1559--1573.}

\noindent{$\bullet$ Emanuel KA.} {1994}.
{Atmospheric convection.
Pp.1--580.
Oxford University Press: 
New York and Oxford.}

\noindent{$\bullet$ Ertel H.} {1942}.
{Ein neuer hydrodynamischer Wirbelsatz.  
{\it Meteorologische Zeitschrift.}
{\bf 59} (9):
277--281.}

\noindent{$\bullet$ Gao S-T, Zhou F-F.} {2008}.
{Water vapour potential vorticity and its applications in tropical cyclones.
{\it Chin. Phys. Lett.}
{\bf 25} (10):
3830--3833.}

\noindent{$\bullet$ Hauf T, H\"{o}ller H.} {1987}.
{Entropy and potential temperature.
{\it J. Atmos. Sci.}
{\bf 44} (20):
2887--2901.}

\noindent{$\bullet$ Hoskins BJ, McIntyre ME, Robertson AW} {1985}.
{On the use and significance of isentropic potential vorticity maps.
{\it Q. J. R. Meteorol. Soc.}
{\bf 111} (470):
877--946. (H85)}

\noindent{$\bullet$ Marquet P.} {2011}.
{Definition of a moist entropic potential temperature. Application to FIRE-I data flights.
{\it Q. J. R. Meteorol. Soc.}
{\bf 137} (656):
768--791. (M11).
\url{http://arxiv.org/abs/1401.1097}.
{\tt arXiv:1401.1097 [ao-ph]}}

\noindent{$\bullet$ Marquet P, Geleyn J-F.} {2013}.
{On a general definition of the squared Brunt-V\"{a}is\"{a}l\"{a} frequency associated with the specific 
moist entropy potential temperature.
{\it Q. J. R. Meteorol. Soc.}
{\bf 139} (670):
85--100.
\url{http://arxiv.org/abs/1401.2379}.
{\tt arXiv:1401.2379 [ao-ph]}
and
\url{http://arxiv.org/abs/1401.2383}.
{\tt arXiv:1401.2383 [ao-ph]}}

\noindent{$\bullet$ Rivas Soriano LJ,  Garc\'{\i}a D\'{\i}ez EL.} {1997}.
{Effect of ice on the generation of a generalized potential vorticity.
{\it J. Atmos. Sci.}
{\bf 54} (10):
1385--1387.}

\noindent{$\bullet$ Santurette P, Joly A.} {2002}.
{ANASYG/PRESYG, M\'et\'eo-France new graphical summary of the synoptic situation.
{\it Meteorol. Appl.}
{\bf 9}:
129--154.}

\noindent{$\bullet$ Schubert W, Hausman SA, Garcia M, Ooyama KV, Kuo H-C.} {2001}.
{Potential vorticity in a moist atmosphere.
{\it J. Atmos. Sci.}
{\bf 58} (21):
3148--3157. (S01)}

\noindent{$\bullet$ Schubert W.} {2004}.
{A generalization of Ertel's potential vorticity to a cloudy, precipitating atmosphere.
{\it Meteorologische Zeitschrift.}
{\bf 13} (6):
465--471.}

\noindent{$\bullet$ Schubert W, Ruprecht E, Hertenstein R, Nieto-Ferreira R, 
Taft R, Rozoff C, Ciesielski P, Kuo H-C.} {2004}.
{English translations of twenty-one of Ertel's papers 
on geophysical fluid dynamics.  
{\it Meteorologische Zeitschrift.}
{\bf 13} (6):
527--576.}

\noindent{$\bullet$ Shapiro MA, Keyser D.} {1990}.
{Fronts, jet streams and the tropopause.
Chapter 10, p.167--191, in 
{\it Extratropical cyclones: the Erik Palm\'en memorial volume},
by C.W. Newton and E.O. Holopainen.
American Meteorological Society.}

\noindent{$\bullet$ Vallis G.K.} {1992}.
{Mechanisms and parameterizations of geostrophic adjustment 
and a variational approach to balanced flow.
{\it J. Atmos. Sci.}
{\bf 49} (13):
1144--1160.}

%\end{thebibliography}

\end{document}